\documentclass[10pt,conference]{IEEEtran}
\usepackage{url}
\usepackage{makecell}
\AtBeginDocument{%
  \providecommand\BibTeX{{%
    \normalfont B\kern-0.5em{\scshape i\kern-0.25em b}\kern-0.8em\TeX}}}

\usepackage{cite}
\usepackage{url}
\usepackage{color}
\usepackage{xspace}
\usepackage{xcolor}
\usepackage{booktabs}
\usepackage{graphicx}
\usepackage{textcomp}
\usepackage{enumerate}
\usepackage{amsmath,amssymb,amsfonts}
\usepackage{subfigure}
\usepackage{algorithm}
\usepackage{multirow}
\usepackage{amsmath}
\usepackage{algorithmicx} 
\usepackage{algpseudocode}
\usepackage{verbatim}
\usepackage{makecell}

\usepackage{setspace}

\begin{document}
%%
%% The "title" command has an optional parameter,
%% allowing the author to define a "short title" to be used in page headers.
\title{Graph-based Fuzz Testing for Deep Learning Inference Engines}
% \author{[Anonymous Author(s)]}

\author{
\IEEEauthorblockN{Weisi Luo\IEEEauthorrefmark{2}, Dong Chai\IEEEauthorrefmark{2}, Xiaoyue Run\IEEEauthorrefmark{2}, Jiang Wang\IEEEauthorrefmark{2}, Chunrong Fang\IEEEauthorrefmark{1}, Zhenyu Chen\IEEEauthorrefmark{1}} 
  
\IEEEauthorblockA{\IEEEauthorrefmark{2}HiSilicon, Huawei, China} 
\IEEEauthorblockA{\IEEEauthorrefmark{1}State Key Laboratory for Novel Software Technology, Nanjing University, China} 
Corresponding author: zychen@nju.edu.cn
}

%%
%% The "author" command and its associated commands are used to define
%% the authors and their affiliations.
%% Of note is the shared affiliation of the first two authors, and the
%% "authornote" and "authornotemark" commands
%% used to denote shared contribution to the research.

%%
%% By default, the full list of authors will be used in the page
%% headers. Often, this list is too long, and will overlap
%% other information printed in the page headers. This command allows
%% the author to define a more concise list
%% of authors' names for this purpose.
% \renewcommand{\shortauthors}{Trovato and Tobin, et al.}

%%
%% The abstract is a short summary of the work to be presented in the
%% article.
\maketitle
\begin{abstract}
With the wide use of Deep Learning (DL) systems, academy and industry begin to pay attention to their quality. Testing is one of the major methods of quality assurance. However, existing testing techniques focus on the quality of DL models but lacks attention to the core underlying inference engines (i.e., frameworks and libraries). Inspired by the success stories of fuzz testing, we design a graph-based fuzz testing method to improve the quality of DL inference engines. This method is naturally followed by the graph structure of DL models.
A novel operator-level coverage criterion based on graph theory is introduced and six different mutations are implemented to generate diversified DL models by exploring combinations of model structures, parameters, and data inputs. The Monte Carlo Tree Search (MCTS) is used to drive DL model generation without a training process. The experimental results show that the MCTS outperforms the random method in boosting operator-level coverage and detecting exceptions. Our method has discovered more than 40 different exceptions in three types of undesired behaviors: model conversion failure, inference failure, output comparison failure. The mutation strategies are useful to generate new valid test inputs, by up to an 8.2\% more operator-level coverage on average and 8.6 more exceptions captured. 
\end{abstract}


%% A "teaser" image appears between the author and affiliation
%% information and the body of the document, and typically spans the
%% page.
%\begin{teaserfigure}
%  \includegraphics[width=\textwidth]{sampleteaser}
%  \caption{Seattle Mariners at Spring Training, 2010.}
%  \Description{Enjoying the baseball game from the third-base
%  seats. Ichiro Suzuki preparing to bat.}
%  \label{fig:teaser}
%\end{teaserfigure}

%%
%% This command processes the author and affiliation and title
%% information and builds the first part of the formatted document.
% \maketitle

\section{Introduction}
Deep Learning (DL) is a popular method for hard computational problems in various domains, such as image classification\cite{r71} and speech recognition\cite{r72}. There are almost innumerable combinations of DL frameworks, data sets, models, platforms, and so on \cite{r60}. On the hardware side, the platforms are tremendously diversified, ranging from common processors (e.g., CPUs, GPUs and NPUs ) to FPGAs, ASICs, and exotic accelerators such as analog and mixed-signal processors. These platforms come with hardware-specific features and constraints that enable or disrupt inference depending on DL models and scenarios.
On the software side, a number of DL inference engines invoked by DL applications commonly serve for optimizing various DL models and performing run-time acceleration inference on the above devices, such as NVIDIA TensorRT\cite{r61}, TensorFlow-Lite\cite{r62} and Alibaba MNN\cite{r55}. 
Hence, the quality of DL inference engines supporting DL models is important to the quality of applications. To the best of our knowledge, there is still a lack of systematic testing methods for DL inference engines. 

Fuzz testing is a widely used automated testing technique that generates random data as inputs to detect crashes, memory leaks, and other failures in software \cite{r63}. Fuzz testing has been shown to be a promising direction for quality assurance of DL systems \cite{r46} \cite{r59}. However, the existing fuzz testing techniques depend heavily on manually designed DL models, and usually perform perturbations, e.g., layer addition, layer removal, data shuffle, noise perturbation on such existing models or data \cite{r1} \cite{r2} \cite{r9}. Different from fuzz testing on DL models, fuzz testing on DL inference engines is expected to generate diversified DL models by exploring combinations of model structures, parameters, weights and inputs. It is 
very challenging to generate such complex graph structured data automatically and effectively.
 
The first challenge in fuzz testing of DL inference engines is to generate diversified DL models to trigger different structured parts of a given DL inference engine. These structured parts include model structure conversion, model optimizations (e.g., operator fusion), data format conversion (e.g., NCHW, NHWC, NC4HW4\cite{r55}), operator implementation, calculation graph scheduling, 
%data-movement a hierarchy of tiles
data-movement a hierarchy of tiles, and so on \cite{r3}. The second challenge is to capture the behaviors of each test, such that fuzz testing can be well directed in generating new tests. The existing neural coverage criteria 
of DL models cannot work in this testing scenario, because the inputs for testing DL inference engines are DL models.
A novel criterion is required to capture behaviors of DL inference engines rather than DL models. It is natural to design testing methods inspired by the graph structure of DL models and analyze the behaviors of the DL inference engine under test against different DL models.

In this paper, a novel graph-based fuzz testing method is designed to test DL inference engines via generating DL models as digraphs. 
The idea of this method naturally conforms to the graphic structure of the DL model. 
It alleviates the problem of generating a large number of diverse DL models. 
Since DL models are not a simple digraph, they have rich characteristics of deep learning elements. 
Thus, beyond the basis of DL model generation, four model-level mutations, including graph edges addition, graph edges removal, block nodes addition, block nodes removal, and two source-level mutations, including tensor shape mutation and parameter mutation, are proposed to generate more diversified DL models effectively. 

To guide more effective fuzz testing for a given DL inference engine under test, the dynamic behaviors of each test (a DL model) should be captured. The operator-level coverage criterion is introduced from graph theory to measure the parts of a DL inference engine's logic exercised by a test set (a number of DL models) based on operator types, structures (e.g., input degree, output degree and single edges from graph theory), tensor shapes and parameters. 
The success ratio and operator-level coverage of an operator are used as feedback to the Monte Carlo Tree Search (MCTS). MCTS is used to solve the search problem to determine whether an operator is select or not in a new model, such that the most promising blocks can be chosen to generate stochastic DL models.

The experiments have been designed and conducted with MNN \cite{r55} on X86 CPU and ARM CPU. 
%这里重新组织下，不要跟摘要完全相同
%The experimental results show that the MCTS outperforms random method in boosting operator-level coverage and detecting exceptions. 
%Our graph-based fuzz testing method have discovered more than 40 different exceptions in three types of undesired behaviors, including models conversion failures, inference failures, output %comparison failures. The mutation strategies are useful to generate new valid test inputs, by up to a 8.2\% more operator-level coverage on average and 8.6 more exceptions captured. 
The experimental results shows that our method is effective in detecting exceptions of DL inference engine, with more than 40 different exceptions discovered, such as crashes, inconsistencies, nan and inf bugs.
Besides, the MCTS-based search performs better than the random-based search in boosting operator-level coverage(6.7\% more) and detecting exceptions (9.7 more).
Furthermore, the mutation strategy helps produce 8.2\% more operator coverage and 8.6 more exceptions detected on average.
Our main contributions in this paper are as follows.
\begin{itemize}
\item A novel graph-based fuzz testing technique is proposed for DL inference engines, where DL models are defined as digraphs from a natural and effective basis.

\item A novel operator-level coverage criterion is introduced to enhance fuzz testing via a reward-guided metric, which can estimate the amount of DL inference engine's logic explored. 

\item Some graph-based model-level mutations (graph edges addition, graph edges removal, block nodes addition, block nodes removal ) and source-level mutations (tensor shape mutation, parameter mutation) are proposed to generate diversified DL models.

\item An experimental evaluation on an industrial inference engine is conducted. The results show that the operator-level coverage guided testing framework improves the effectiveness in detecting exceptions.

% \item We demonstrate that the problem of finding a series of blocks that would result in detecting more exceptions while maximizing operator-level coverage can be formulated as a search problem. We present a MCTS-based search algorithm for solving this problem efficiently.

% show that MCTS-based search algorithm can be used to decision processes to improve operator-level coverage by up to a XXX\% more operator-level coverage on average and XXX more exceptions captured.

% \item We conduct an experimental evaluation and the results show that subgraphs defined in block corpus improve the effectiveness in detecting  exceptions of specific structures.
% %through generating more targeted inputs with specific topologies.
\end{itemize}

More   details   of   graph-based   fuzz   testing   and   the   experiments  can  be  found  at  \url{https://github.com/gbftdlie/Graph-based-fuzz-testing}.

\section{Background}

\subsection{Workflow of DL inference engines}

\begin{figure}[ht]
  \centering
   \includegraphics[width=0.42\textwidth]{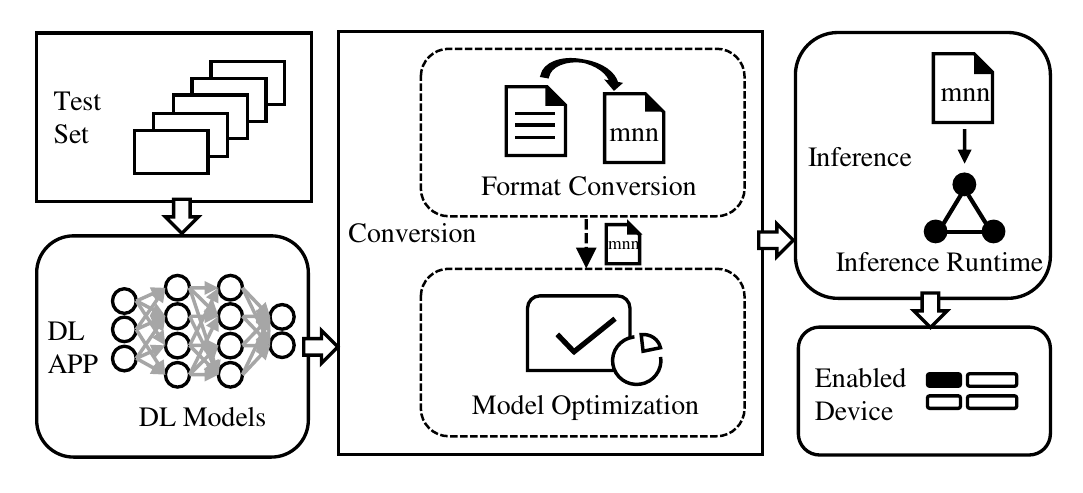}
  \caption{Inference workflow with MNN}
%   \Description{}
  \label{p1}
\end{figure}

DL inference engines are invoked by DL applications to load and run models on devices with inference hardware accelerators.
The workflows of existing inference engines are similar. Take MNN \cite{r55}(a lightweight DL inference engine developed by Alibaba Inc) as an example,
%For example, MNN\cite{r55} is a lightweight deep neural network inference engine. It can be divided into two parts: Converter and Interpreter. 
%MNN\cite{r55} is a typical lightweight DL inference engine a
as shown in Fig. \ref{p1}, the inference workflow can be roughly divided into two phases: 
(1) Conversion: phase of converting those training framework models (e.g., TensorFlow (Lite), Caffe, and ONNX) into MNN models and optimizing DL models by operator fusion, operator substitution, and layout adjustment. Furthermore, MNN models can be quantized optionally.
(2) Inference: phase of loading MNN model and inferring. The interpreter of MNN consists of engine and backends. The former is responsible for loading a MNN model and scheduling a computational graph; the latter includes the memory allocation and the operator implementation under each computing device. 

% DL inference engine enables a trained model to run on devices with inference hardware accelerators using CPUs, GPUs, DSPs, NPUs, etc. 
% The converter is responsible for converting a model trained via some DL framework to a MNN model. The converter also provides  optimization
% customized for the DL model, such as operator fusion, operator substitution, and layout adjustment.
% Finally, the interpreter load the MNN model and do on-device inference\cite{r55}. 

\subsection{Limitations of existing testing techniques}
Fuzz testing has been proved to be effective in exploring the input space of DL system testing. An important part of fuzz testing is how to feedback.
% A coverage criterion is used to measure what percent of the potential behaviors could be tested by a given set of inputs.
% \fcr{Citations for neural coverage} 
%For this purpose, a coverage criterion divides the input space for a given system into equivalence classes and calculates how many of the equivalence classes have at least one instance input in a given set of inputs. When an equivalence class has at least one instance input in a given set of inputs, the equivalence class is said to be covered. 
DeepXplore\cite{r2} introduced the concept of neuron coverage for measuring the parts of a DL system's logic exercised by a set of test inputs based on the number of neurons activated (i.e., the output values are higher than a threshold) by the inputs. 
Enlightened by DeepXplore, a number of different coverage criteria have been applied to DL testing, such as TensorFuzz\cite{r1}, Deeptest\cite{r66} and Deepgauge\cite{r35}. 
These testing techniques focus on testing the quality of specific DL models. 
However, in DL inference engine testing, when the input model is changed, the coverage will be invalid as feedback.

Another important part of fuzz testing is to generate test inputs via mutating a given set of inputs.
DeepMutation \cite{r9} proposed a set of mutation operators, including changing weights, biases and inputs via disturbing, exchanging neurons within a layer, adding or removing a layer of which input and output dimensions are equal, like batch normalization, etc. This method mutates a single model to simulate error scenarios. TensorFuzz is another fuzz test generation tool based on input data mutation for a DL model. These methods can also effectively guarantee the quality of specific DL models. But they cannot generate a large number of diversified models for testing inference engines, that is, despite a large amount of input generated, the given model has not changed or changed very little. Therefore they are unlikely to detect erroneous behaviors and trigger different parts of a DL inference engine's logic. 

In general, 
existing testing techniques focus on the quality of DL models, but lack attention to testing DL inference engines. Diversified combinations of operators (models) are more capable of triggering DL inference engine issues than an specific combination of operators (a specific/single model). 
The issues triggered by a single model are limited. We need to generate a large number of models by combining operators as the test inputs of DL inference engines .

%existing testing methods are not effectively for DL inference engines. They cannot generate a large number of mutated models and capture the problems in inference engines and generate effective feedback.

% mutation testing (Deepmutation \cite{r9}). 
% The  main  restriction  of these  approaches  is  that  it  limit  its  perturbations  (e.g.,  layer addition,  layer  removal,  data  shuffle,  noise  perturbation)  to tiny  changes  on  existing  models  or  data. 
% Further, DL inference engines or frameworks have made optimizations responding to demands for faster inference. CULASS \cite{r3} of CUDA provides specialized data-movement and a hierarchy of tiles to improve performance. TensorFlow \cite{r7} \cite{r10} \cite{r11} applies graph optimization to improve the performance of computations. MNN\cite{r55} applies perform optimization in the graph level or operator fusion and replacement.
% Test inputs for these testing scenarios require various operators and various structures in DL models. Existing test techniques above are difficult to provide such test inputs for DL inference engine.
% Thereby these optimizations also challenge fuzz testing of DL inference engine to generate DNNs with various input shapes and topologies to assure correctness.

\section{Methodology}
In this section, we provide detailed technical descriptions of graph-based fuzz testing for DL inference engines. 
%First, we define and explain the concepts of digraph, subgraph, block, block corpus, %Operator-level coverage, etc. Next, we provide a black box fuzz testing framework for DL inference %engine. Finally, we describe block corpus, block chooser, mutations and  shapes\&parameters %calculator in our approach. 

\subsection{Definitions} \label{definitions}
The stochastic networks generation involves the following definitions: \textbf{digraph}, \textbf{subgraph}, \textbf{block}, \textbf{block corpus}, \textbf{mutation action}.

\textbf{Digraph}. Graph theory provides an excellent framework for studying and interpreting neural network topologies \cite{r13}. A neural network can be denoted by a digraph ${G}=({V},{E})$. V is a set of operators (e.g., Conv2d, Relu and Softmax).  $E \subseteq \left\{ (x,y)|(x,y)\in V^2  \bigwedge x \neq y \right\} $ is a set of directed edges which are ordered pairs of distinct operators (e.g., ${x}$ and ${y}$).
% (i.e., an edge is associated with two distinct operators, ${x}$ and ${y}$ ) . 
In neural networks, edges are data flows. 

\textbf{Subgraph}. From the introduction above, some specified structures of DL models will be specially processed (e.g., operator fusion and replacement\cite{r64}\cite{r65} to run a faster inference). There is a very low probability that these specified structures could be generated randomly. %by stochastic network generator.
 Thus subgraphs are applied to blocks to define those specified structures directly in testing. 
Formally, digraph ${{G}'} = ({{V}'}, {{E}'})$  is a subgraph of ${G}$ iff ${{V}'}\subseteq{V},{{E}'}\subseteq{E} \bigwedge (({x},{y})\in{{E}'}\rightarrow {x},{y}\in{{V}'})$ . ${x}$ and ${y}$ are two distinct operators.

\begin{figure}[ht]
  \centering
  \includegraphics[width=0.6\linewidth]{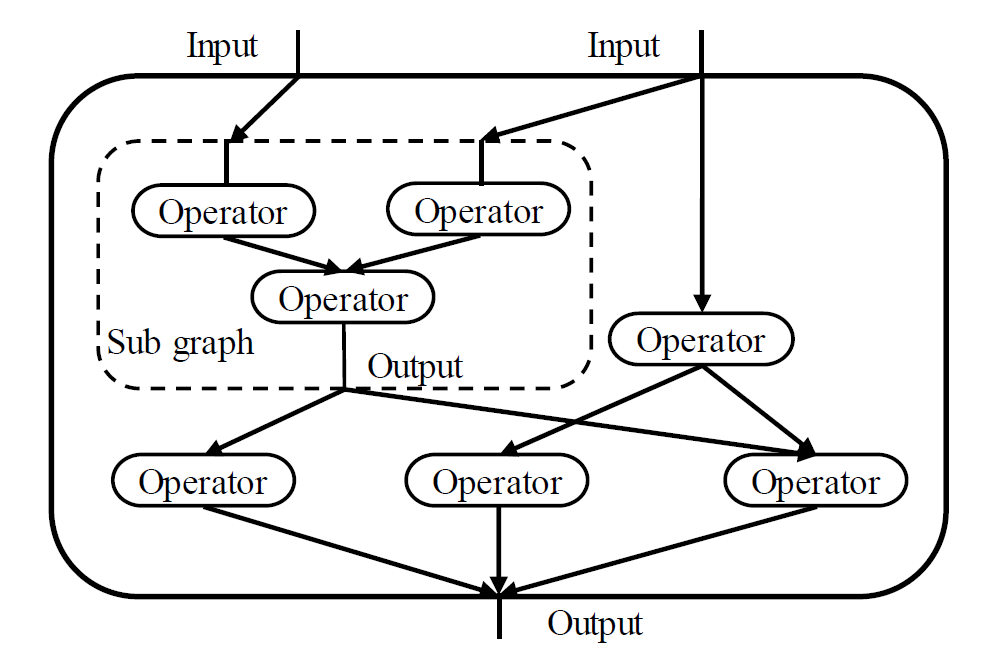}
  \caption{ A network is constructed by operators and subgraphs.}
%   \Description{}
  \label{p_subgraph}
\end{figure}

\textbf{Block}. Subgraphs or operators of a neural network are defined as blocks in this paper. A network is constructed by operators and subgraphs as shown in Fig. \ref{p_subgraph}.

\textbf{Block corpus}. A block corpus contains blocks to be chosen and their attributes, including block name, allowed range of in-degree and out-degree, inner edges of the block. Inner edges are required when the block is a subgraph and can be empty otherwise. 
To construct the block corpus, the tester need to first confirm the types of operators and subgraphs to be tested, and then fill in their above attributes into the block corpus.

\textbf{Mutation action}. 
Let ${I_1},{I_2},$\textellipsis$,{I_n}$ be a sequence of mutated test sets, where ${I_k}$ is generated by ${k}$-th mutation action $MA(bs_k,ms_k)$. ${bs_k}$ and ${ms_k}$ are the blocks selection and mutation operators selection in ${k}$-th mutation action respectively. A tuple of the two actions forms a complete action ${MA}({bs},{ms})$.

\subsection{Operator-level Coverage Criterion}
% \textbf{Operator-level Coverage}.
%加个引文
Traditional code coverage criteria are ineffective in DL testing. As discussed in Section II.B, recently proposed neuron coverage criteria are still invalid, as DL inference engine testing involves different models.
%criterion 是单数， criteria是复数
A novel operator-level criterion is proposed to capture differences in models combined by operators, which provide feedback to guide the proposed graph-based fuzz testing to generate diversified models. 
As defined in \ref{definitions}, we use the model structures, input tensor shapes, parameters to characterize behaviors of operators in DL models. 
As defined in \ref{definitions}, we use the input degree (the number of input data flows), the output data flows (the number of output data flows), input tensor shapes (NHWC, etc.) and parameters (e.g., dilation of Conv2d) of operators to characterize behaviors of operators in DL models.

% at operator-level. 
%If a new test sample produces additional coverage, it will be added in input test set.
Given a block corpus ${BC}$ and a test set ${I}$, operator-level coverage criterion is defined as follows: 

\textbf{ Operator Type Coverage(OTC)}.
Let ${n_{t}}$ be the number of total types of operators defined in ${BC}$. 
Let $OTC\_op({c},{I})$ be 1 when the operator ${c}$ is in the ${I}$, and be 0 otherwise. 
The OTC of  ${I}$ is defined as :
% $${{OTC}({I})}=\frac{\sum{OTC\_op({c},{I})}}{n_{t}}$$ 
\begin{equation}
{{OTC}({I})}=\frac{\sum{OTC\_op({c},{I})}}{n_{t}}\label{eq_OTC}
\end{equation}

% $${{OTC}({I})}=({f_{t}} ({I})/{n_{t})\times100\%}$$ .

% Let ${f_{t} ({I})}$ be the number of types of the operator in  ${I}$.

% Let $OTC\_op({c},{I})$ be 1 when the operator ${c}$ is in the ${I}$, and be 0 otherwise.  It is defined as the ratio of the number of types for a test set ${I}$ . 
% The OTC of  ${I}$ is defined as : $${{OTC}({I})}=\frac{{f_{t}} ({I})}{n_{t}}$$ 
% % $${{OTC}({I})}=({f_{t}} ({I})/{n_{t})\times100\%}$$ . 
\textbf{Input Degree Coverage(IDC)}. Let $n_{id\_c}$ be the total number of different input degrees of operator ${c}$ in ${BC}$. Let $f_{id\_op}({I})$ be the number of different input degrees for operator $c$ in  $I$. The input degree coverage of operator $c$ is defined as the ratio of $f_{id\_op} (I)$ to $n_{id\_op}$: 
% ${IDC\_op}({c},{I})=({f}_{id\_op} (I)/{n}_{id\_c})\times100\%$. 
${IDC\_op}({c},{I})=\frac{{f}_{id\_op} (I)}{{n}_{id\_c}}$. 
% The IDC of  ${I}$ is defined as: $${IDC(I)}=\frac{\sum{IDC\_op}({c},{I})}{{n}_{t}}$$
\begin{equation}
{IDC(I)}=\frac{\sum{IDC\_op}({c},{I})}{{n}_{t}}\label{eq_IDC}
\end{equation}

% $${IDC(I)}=(\sum{IDC\_op}({c},{I})/{n}_{t})\times100\%$$. 

\textbf{Output Degree Coverage(ODC)}.
 Let ${n_{od\_c}}$ be the total number of different output degrees of operator ${c}$ in ${BC}$. Let ${f_{od\_op}}$ ({I}) be the number of different output degrees of operator ${c}$ in  ${I}$. The output degree coverage of operator ${c}$ is defined as the ratio of ${f_{od\_op}} ({I})$ to ${n_{od\_c}}$: 
 ${ODC\_op({c},{I})}=\frac{{f_{od\_op}} (I)}{n_{od\_c}}$.
%  ${ODC\_op({c},{I})}=({f_{od\_op}} (I)/{n_{od\_c}})\times100\%$.
 The ODC of  ${I}$ is defined as: 
%  $${ODC(I)}=\frac{\sum ODC({c},{I})}{n_{t}}$$
 \begin{equation}
{ODC(I)}=\frac{\sum ODC\_op({c},{I})}{n_{t}}\label{eq_ODC}
\end{equation}
%  $${ODC(I)=(\sum ODC({c},{I}) /n_{t})\times100\%}$$.

\textbf{Single Edge Coverage (SEC)}.
 Let ${f_{se} (c,I)}$ be the number of different edges that directed from the operator ${c}$ to others in  ${I}$. The number of total edges that directed from the operator ${c}$ to others in ${BC}$ is ${n_{t}}$. The single edge coverage of operator ${c}$ is defined as the ratio of ${f_{se} ({c},{I})}$  to ${n_{t}}$:  
 ${SEC\_op({c},{I})}=\frac{f_{se} ({c},{I})}{n_{t}}$. 
%  $${SEC\_op({c},{I})=(f_{se} ({c},{I})/n_{t}\times100\%}$$. 
 The SEC of ${I}$ is defined as:
%  $${SEC(I)}=\frac{\sum SEC\_op({c},{I}) }{n_{t}}$$   
\begin{equation}
{SEC(I)}=\frac{\sum SEC\_op({c},{I}) }{n_{t}}\label{eq_SEC}
\end{equation}
 
%  $${SEC(I)=(\sum SEC({c},{I}) /n_{t})\times100\%}$$.  

% \textbf{Multiple Edges Coverage(MEC)}.
%  Let ${n_{me}}$ be the number of different multi-input operators defined in ${BC}$. Let ${MEC\_op({c},{I})}$ be 1 when the multi-input operator ${c}$ in test set ${I}$ has same inputs, and be 0 otherwise. The MEC of test set ${I}$ is defined as: 
%  $$ {MEC(I)}=\frac{\sum MEC\_op({c},{I}) }{n_{me}}$$  
% %  $$ {MEC(I)=(\sum MEC\_op({c},{I}) /n_{me}) \times100\%}$$. 

\textbf{Shapes$\&$Parameters Coverage(SPC)}. 
 Let ${n_{maxspc}}$ be the expected maximum of shape$\&$parameter. Let ${f_{spc} ({c},{I})}$  be the number of distinct vectors including tensor shapes and parameters for operator ${c}$ in  ${I}$. The SPC of operator c in ${I}$ is defined as: 
$ {SPC\_op({c},{I})} = \frac{f_{spc}({c},{I})} {n_{maxspc} }$. 
% $ {SPC\_op({c},{I}) = f_{spc}({c},{I})/ n_{t} }$ . 
The SPC of  ${I}$ is defined as: 
% $${SPC({I})}=\frac{(\sum SPC\_op({c},{I}) / n_{maxspc})}{n_t}$$
\begin{equation}
{SPC({I})}=\frac{\sum SPC\_op({c},{I})}{n_t}\label{eq_SPC}
\end{equation}

% $${SPC({I})=((\sum SPC\_op({c},{I}) / n\_to)/n_spc)\times100\%}$$. 

\textbf{Operator-level Coverage (OLC)}. 
%  Let ${m(c)}$ be the number of metrics involved for operator ${c}$.
% The OLC of operator ${c}$ of test set ${I}$ is:
% $OlC\_op({c},{I}) =(IDC\_op({c},{I}) +ODC\_op({c},{I}) +SEC\_op({c},{I}) +MEC\_op({c},{I})
% +SPC\_op({c},{I}))/{m(c)})\times100\% $ , when ${c}$ is a  multi-input operator, and $MEC\_op({c},{I})$ does not involve otherwise.
% Let ${m_I}$ be the number of metrics of ${I}$. OLC of  ${I}$ is defined as: 
Let  OLC of the operator ${c}$ be the weighted mean of a set of metrics $ Z\_op = \{{OTC\_op({c},{I})} ,  {IDC\_op({c},{I})} ,   {ODC\_op({c},{I})} ,  {SEC\_op({c},{I})} ,
%  \\{MEC\_op(I)},
 \\{SPC\_op({I})} \}$ with corresponding non-negative weights  $\{{w\_op_1},{w\_op_2},...,{w\_op_5}\}$. OLC of  operator ${c}$ is defined as: 
% $${OLC\_op({c},{I}) = \frac{\sum{w\_op_i}{m\_op_i}}{\sum{w_i}} ,   {m\_op_i}\in Z\_op}$$.
\begin{equation}
{OLC\_op({c},{I}) = \frac{\sum{w\_op_i} {m\_op_i}}{\sum{w\_op_i}} ,   {m\_op_i}\in Z\_op}\label{eq_OLCOP}
\end{equation}

% $$OLC\_op({c},{I}) = \frac{\sum{w_i}{m\_op_i}}{\sum{w_i}} ,  {m\_op_i}\in Z\_op , \sum{w_i} = 1$$

Let OLC of the test set ${I}$ be the weighted mean of a set of metrics $ Z = \{{OTC({I})} ,  {IDC(I)} ,  {ODC(I)} , {SEC(I)} , 
% \\{MEC(I)}, 
\\{SPC({I})} \}$ with corresponding non-negative weights $\{{w_1},{w_2},...,{w_5}\}$. Formally,  OLC of  test set ${I}$ is defined as: 
% $${OLC({I}) = \frac{\sum{w_i}{m_i}}{\sum{w_i}} ,   {m_i}\in Z }$$
\begin{equation}
{OLC({I}) = \frac{\sum{w_i}{m_i}}{\sum{w_i}} , {m_i}\in Z }\label{eq_OLC}
\end{equation}

% $$OLC({I}) = \frac{\sum{w_i}{m_i}}{\sum{w_i}} ,  {m_i}\in Z , \sum{w_i} = 1$$
% The values of weights depend on the sensitivity of DL inference engine to the metrics above. 
Some weights may be zero. For example, the weight of $ODC({I})$ should be 0 when expected output degree of operators are all 1 in test samples of test set ${I}$. 
%  Let ${m(c)}$ be the number of metrics involved for operator ${c}$.
% The OLC of operator ${c}$ of test set ${I}$ is:
% $OlC\_op({c},{I}) =(IDC\_op({c},{I}) +ODC\_op({c},{I}) +SEC\_op({c},{I}) +MEC\_op({c},{I})
% +SPC\_op({c},{I}))/{m(c)})\times100\% $ , when ${c}$ is a  multi-input operator, and $MEC\_op({c},{I})$ does not involve otherwise.
% Let ${m_I}$ be the number of metrics of ${I}$. OLC of  ${I}$ is defined as: 

% $OlC({I})=((OTC({I})+IDC({I})+ODC({I})+SEC({I})
% +MEC({I})+SPC({I}))/{m_I})\times100\%$, 

% when test set ${I}$ contains multi-input operator, and $MEC({I})$ is not involved otherwise.

\begin{figure}[ht]
\subfigure[NN1]{
\includegraphics[width=0.14\textwidth]{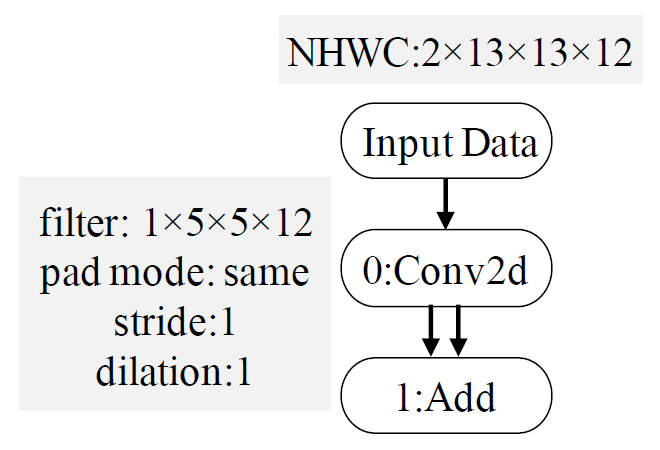}
\label{nn_1}
}
\subfigure[NN2]{
\includegraphics[width=0.08\textwidth]{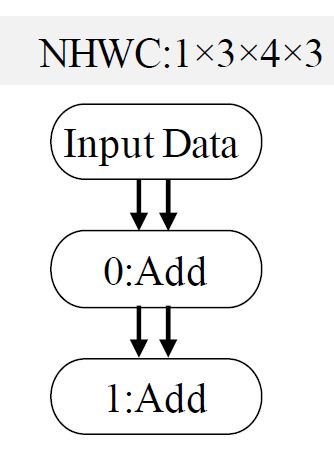}
\label{nn_3}
}
\subfigure[NN3]{
\includegraphics[width=0.2\textwidth]{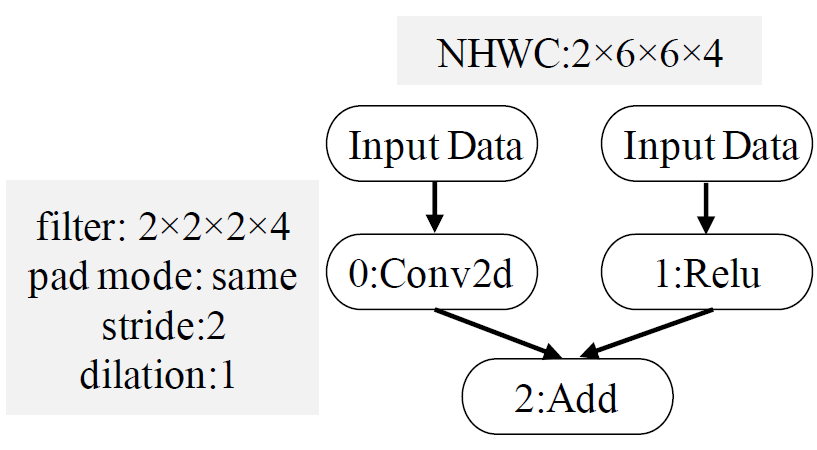}
\label{nn_2}
}
\caption{Three NNs in Test Set {I}}
\label{3nn}
\end{figure}

 For example, Fig. \ref{3nn} shows three NNs in test set ${I}$ generated by block corpus {BC} in TABLE \ref{table1}. Tensor format is NHWC.  Three blocks  Conv2d, Relu and Add, and their input and output degree, are defined. In operator-level coverage, the ${n_{maxspc}}$ of Formula(\ref{eq_SPC}) is set to 10. 
 Operator-level coverage result for each operate and test set ${I}$ are calculated and listed in  TABLE \ref{table2} respectively.

\begin{table}[htbp]
\centering
\caption{Block Corpus of Test Set ${I}$}
\scriptsize
\label{table1}
\begin{tabular}{ c c c c } 
\hline
Block Name&Input Degree&Output Degree&Inner Edges\\
\hline
Conv2d&\{1\}&\{ 0,1,2\}&N/A\\
\hline
Relu&\{1\}&\{ 0,1,2\}&N/A\\
\hline
Add&\{2\}&\{ 0,1,2\}&N/A\\
\hline
\end{tabular}
\end{table}

\begin{table}[htbp]
\centering
\caption{Operator-level Coverage of each operator and test set ${I}$}
\scriptsize
\label{table2}
\begin{tabular}{cccccccc} 
\hline
Object&OTC&IDC&ODC&SEC&SPC&OLC\\
\hline
Conv2d&100\%&100\%&66.7\%&33.3\% &20\% &64\%\\
\hline
Relu&100\%&100\%&33.3\%&33.3\% &10\% &55.3\%\\
\hline
Add&100\%&100\%&66.7\%&33.3\%  &30\% &66\%\\
\hline
${I}$& 100\% & 100\% & 55.6\%&33.3\% &20\%&61.8\%\\
\hline
\end{tabular}
\end{table}

\subsection{Framework}
% The core idea of our approach is to make reward-guided fuzz testing, augmenting tests by mutating blocks and achieving higher operator-level coverage. In generating process, test inputs are modeled as a graph. In reward-guided process, operators are evaluated with their success ratio and coverage changes they induce. 
The core idea of graph-based fuzz testing is to maximize operator-level coverage on a DL inference engine such that as many erroneous behaviors as possible can be exposed. A large number of test samples (i.e., models) can be constructed by mutating existing DL models (modeled as graphs) and operator-level coverage is used as feedback to guide test input generation.

\begin{figure}[ht]
  \centering
  \includegraphics[width=\linewidth]{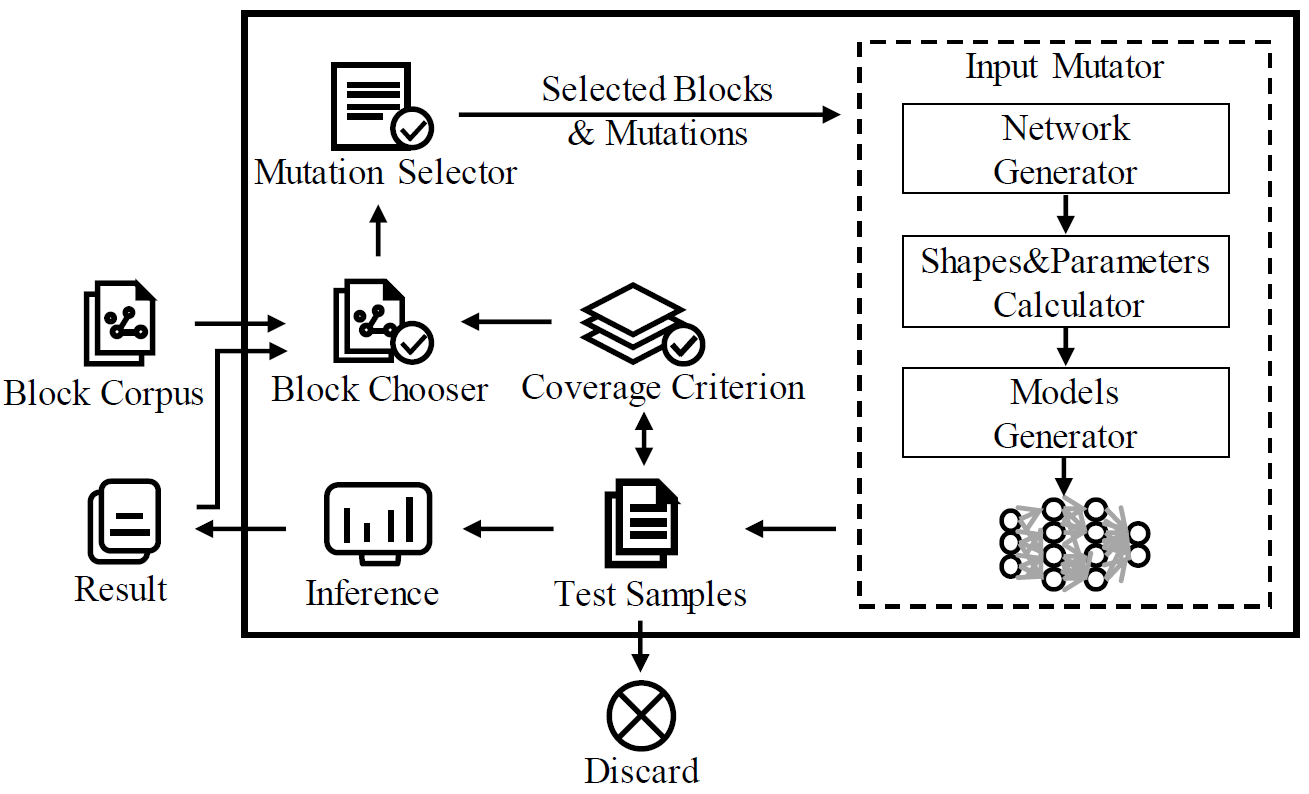}
  \caption{Framework of graph-based testing}
  \label{p2}
\end{figure}

\textbf{Framework}. As depicted in Fig. \ref{p2}, the graph-based fuzz testing framework is composed of block chooser, coverage criterion, input mutator, and mutation selector. 
%block corpus有介绍吗？
For each iteration, the MCTS-based block chooser chooses a set of blocks $b$ from block corpus. The mutation selector chooses one or more mutations scholastically to determine mutating rules ${m}$. Parameters of the mutations are assigned randomly under their constraints. After that, the input mutator determines which actions in ${m}$ will be applied to $b$ where mutating actions (${b}$, ${m}$) are formed and test samples can be generated. The test samples will be run in DL framework (e.g., TensorFlow) whose output data is saved as expected results. The Input Data contains models and their expected results. The coverage criterion takes the mutated samples to check whether current coverage increases. If current coverage increases, the new input data will be added to test set; otherwise, such data will be discarded.
This process runs until reaching a preset threshold, the target number of test samples.

\begin{algorithm}[ht]
\caption{Algorithmic description of our fuzz testing framework} %
\small
\begin{algorithmic}[1]
\Procedure{FuzzWorkflow}{${BC}$, ${M}$, ${tc0}$}
\While{$not$ ${tc0}$}
\State ${b_k}$, ${C}$ = BlockChooser(${R}$, ${tc1}$, ${tc2}$, ${coverage}$)
\State ${source\_m_k}$, ${model\_m_k}$ = MutationSelector (${b_k}$, ${M}$)
\State ${I_k}$ = InputMutation (${b_k}$, ${source\_m_k}$, ${model\_m_k}$)
\State ${coverage_k}$ = CoverageCriterion (${I_k}$)
\If{IsNewCoverage (${coverage_k}$)}
\State ${result_k}$ = MCTSSimulation (${I_k}$)
\State update ${result}$, ${coverage}$, ${I}$
\State MCTSBackpropagation (${C}$, ${R}$, ${result}$, ${coverage}$) 
\EndIf
\EndWhile
\State \Return{$ {result}, {coverage}, {I}$}
\EndProcedure 
\Procedure{InputMutation}{${b_k}$, ${source\_m_k}$, ${model\_m_k}$}
\State ${g}$ = GenerateGraph (${b_k}$, ${model\_m_k}$)
\State ${s}$, ${p}$  = CalcShapesParameter (${g}$, ${source\_m_k}$)
\State ${I_k}$ = GenerateModel (${g}$, ${s}$, ${p}$)
\State \Return{${I_k}$}
\EndProcedure 
\Procedure{BlockChooser}{${R}$, ${tc1}$, ${tc2}$, ${coverage}$}  
\State ${L}$ = MCTSSelection (${R}$, ${tc1}$)
\State ${C}$ = MCTSExpansion (${L}$, ${coverage}$, ${tc2}$) 
\State ${b_k}$  = GetNodesFromPath (${C}$)
\State \Return{${b_k}$, ${C}$}
\EndProcedure 
\end{algorithmic}
\end{algorithm}

\textbf{Algorithm}. The algorithm of our graph-based fuzz testing is shown in Algorithm 1. 
In procedure of FuzzWorkflow, inputs are block corpus (${BC}$), mutations (${M}$), a termination condition (${tc0}$) that is a target number of new inputs. The while loop in Line 2 iterates until ${tc0}$ is reached. In Line 3, blocks are chosen by the block chooser. In Line 4, the mutation selector selects mutations and their parameters. In Line 5, Input Mutation generates test set ${I_k}$ by the blocks and mutations. In Line 6, the operator-level coverage of ${I_k}$ is calculated. In Line 7, ${coverage_k}$ is checked whether it produces additional coverage. In Line 8, MCTS Simulation is made. In Line 9, current test set, results and current operator-level coverage are updated. In Line 10, MCTS Back propagation updates back the result of the inference to update values associated with the nodes on the path from child node ${C}$ to root node ${R}$.

In procedure of InputMutation, inputs are selected blocks ${b_k}$, selected model-level mutations ${model\_m_k}$ and source-level mutations ${source\_m_k}$. In Line 16, graphs are generated from the selected blocks ${b_k}$ and model mutations. In Line 17, input shapes and parameters of each block are generated from the graphs and data mutations. In Line 18, according to the graphs and parameters, the test set ${I_k}$ is generated.

% In procedure of BlockChooser, inputs are the root node ${R}$ of MCTS tree, corpus, termination condition ${tc1}$ that is the maximum levels of the search tree the MCTS can go down, termination condition ${tc2}$ that is the maximum times a MCTS node can be explored. In line 21, choose blocks for InputMutation. In line 22, MCTS Selection is made. And a leaf node ${L}$ is returned. In line 23, MCTS Expansion is applied to  create a new child node ${C}$ of the leaf node ${L}$. The child node ${C}$ could be the lowest coverage operator or a subgraph containing it, and is not chosen in the path before. In line 24-25, index of C and blocks along the path from C to R are returned.  
In the procedure of BlockChooser, inputs are the root node ${R}$ (no operator is set for the root node) of MCTS tree, termination condition ${tc1}$ is the maximum levels of the search tree the MCTS can go down, termination condition ${tc2}$ is the maximum times a  node can be explored. In Line 21, choose blocks for InputMutation. In Line 22, MCTS Selection is made. A leaf node ${L}$ is returned. In Line 23, MCTS Expansion is applied to create a new child node ${C}$ of the leaf node ${L}$. The child node ${C}$ could be the lowest coverage operator or a subgraph containing it, and is not chosen in the path before. In Line 24-25, the index of $C$ and blocks along the path from child node $C$ to root node $R$ are returned.

\subsection{Block Corpus}

\begin{figure}[ht]
\subfigure[Conv2d+Relu+Pow+Concat]{
\includegraphics[width=0.19\textwidth]{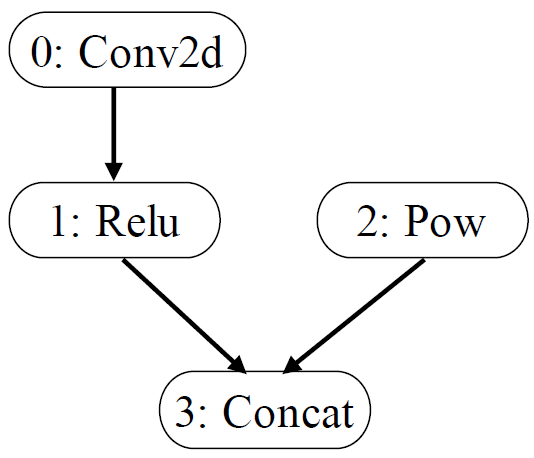}
\label{p3_a}
}
\subfigure[Conv2d+Conv2d+Add+Add]{
\includegraphics[width=0.19\textwidth]{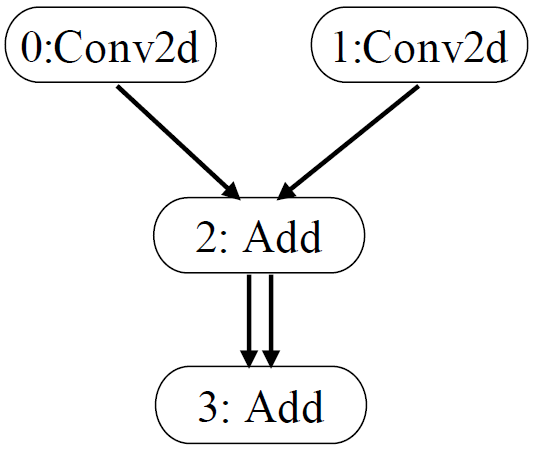}
\label{p3_b}
}

\caption{ Block structures of  Conv2d+Relu+Pow+Concat and  Conv2d+Conv2d+Add+Add}
\label{p3}
\end{figure}

The fuzz testing process maintains a block corpus containing blocks and their attributes, including block name, allowed range of in-degree and out-degree, inner edges of the block. When a block is a subgraph, its block name is defined as the sequence of operators in the subgraph (i.e., block Conv2d+Relu+Pow+Concat in Fig. \ref{p3_a} ) and its inner edges are required. Each element in the adjacency list of inner edges is a pair of source and destination operator index. Taking an operator Conv2d and two subgraphs (shown in Fig. \ref{p3}) for example, Conv2d has exactly one input, and the two subgraphs have two respectively. Allowed range of out-degree of the two are set by test framework, such as \{0,1,2\}. Inner edges of the two subgraphs are \{(0, 1), (1, 3), (2, 3)\} and \{(0, 2), (1, 2), (2, 3) , (2, 3)\} respectively. 
%Conv2d is not involved.

\subsection{Block Chooser}

\begin{figure}[ht]
\subfigure[Selection]{
\includegraphics[width=0.22\textwidth]{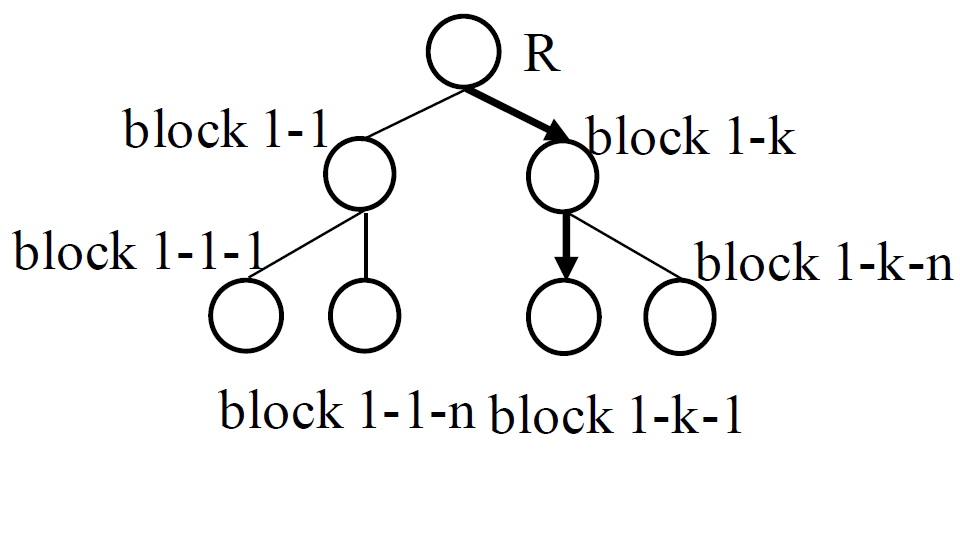}
\label{mctstree_1}
}
\subfigure[Expansion]{
\includegraphics[width=0.22\textwidth]{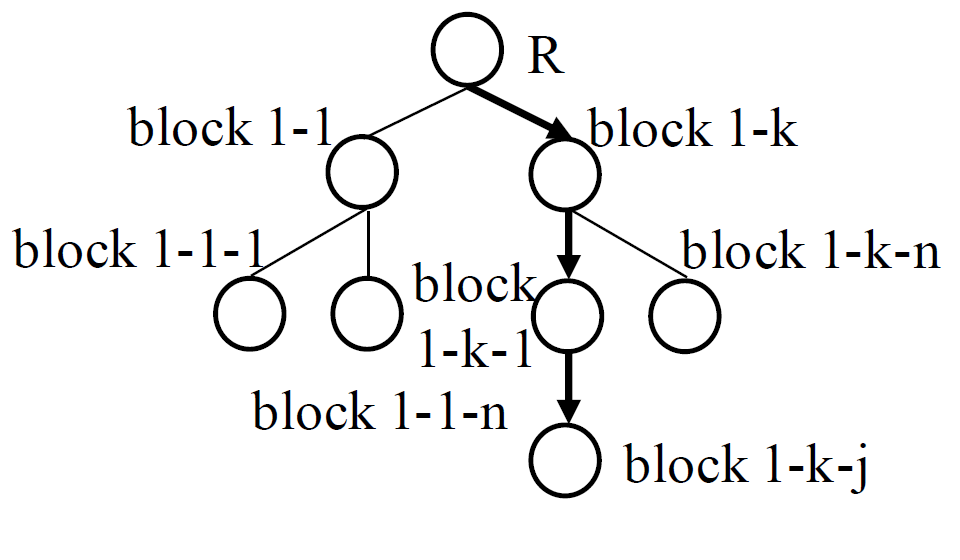}
\label{mctstree_2}
}
\newline
\subfigure[Simulation]{
\includegraphics[width=0.21\textwidth]{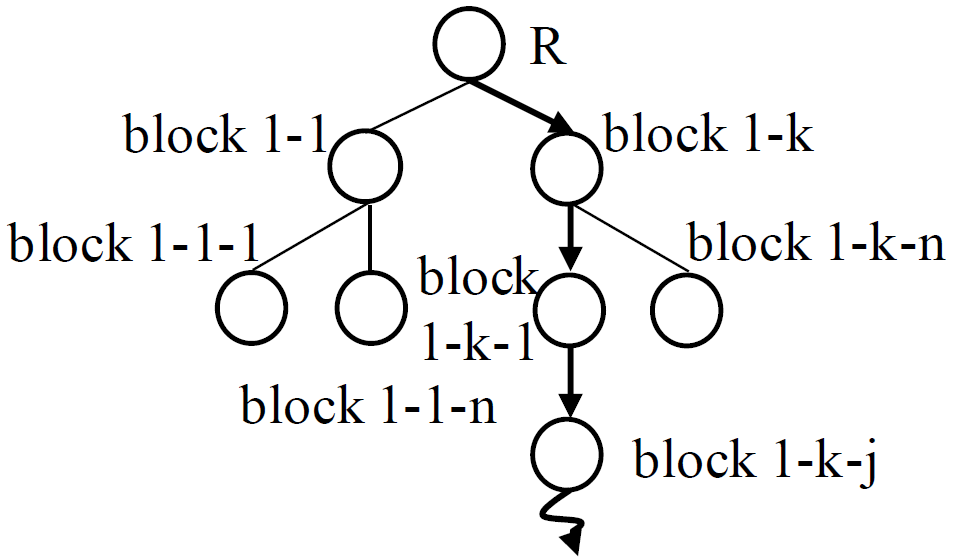}
\label{mctstree_3}
}
\subfigure[Backpropagation]{
\includegraphics[width=0.21\textwidth]{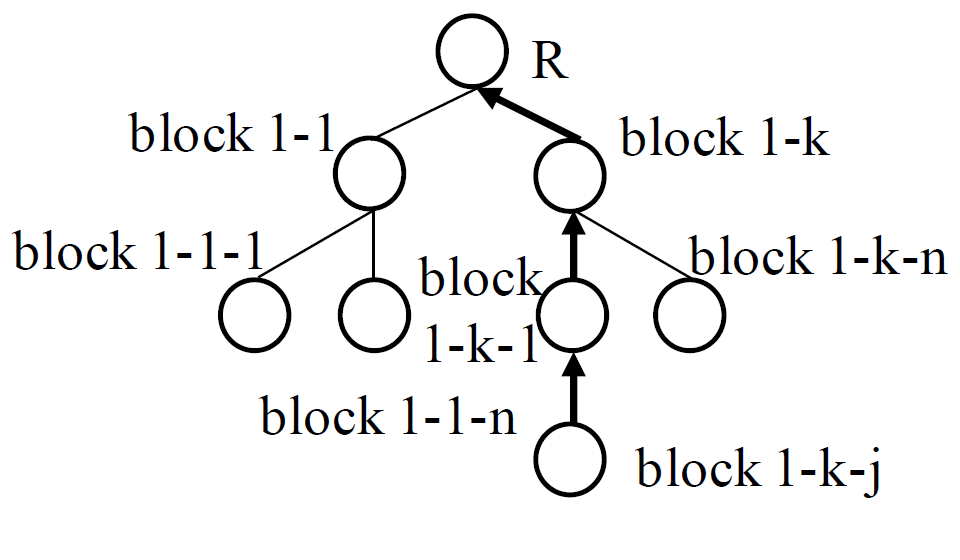}
\label{mctstree_4}
}
\caption{Scheme of a Monte-Carlo Tree Search}
\label{mctstree}
\end{figure}

The block chooser is designed for DL models generation to improve operator-level coverage and suppress duplicated exceptions. In tests that randomly select operators to build DL models, a large number of duplicate exceptions are found.
In block chooser, Monte Carlo Tree Search (MCTS) is used to search the input domain of DL inference engines, such that the most promising blocks can be chosen in block chooser to generate stochastic DL models. Each node of the tree represents an operator in the block corpus.
%On taking an action, one makes a transition from a node to one of its children. 
MCTS dynamically adapts itself to the most promising search regions, where good combinations are likely to find more exceptions. MCTS process shown in Fig. \ref{mctstree} can be divided into the following four steps.

\textbf{Selection}: Starting from the root node $R$, successively select child nodes according to their potentials until a leaf node $L$ is reached. The potential of each child node is calculated by using UCT (Upper Confidence Bound applied to Trees)  \cite{r50} \cite{r51}. UCT is defined as:

\begin{equation}
{potential}=\frac{v}{n}+{e}\times\sqrt{\frac{ln N}{n}} \label{eq1}
\end{equation}

where $v$ refers to the success count of the node, ${n}$ is the visit count of the node, and ${N}$ is the visit count for the parent of the node. ${e}$ is a hyper parameter determining exploration-exploitation trade-off. The maximum levels of the search tree that the MCTS can go down is set as terminal condition 1 (${tc1}$).

\textbf{Expansion}: Unless ${L}$ is a terminal node with the highest potential, create one child node (operator) ${C}$ with the lowest coverage and the operator ${C}$ is not in the path. We pick a operators or a subgraph that contains the operator ${C}$.  

\textbf{Simulation}: Generate stochastic DL models using the blocks in the current path of tree until reaching a terminal condition, and then inference the models. The maximum times a MCTS node can be explored is set as terminal condition 2 (${tc2}$).

\textbf{Back propagation}: Propagates back the result of the inference to update values associated with the nodes on the path from ${C}$ to ${R}$. The path containing the nodes with the highest values in each layer would be the optimal strategy in the test set.

\subsection{Mutatation Selector}

\textbf{Mutations}. The stochastic graphs generated by graph models above, only cover a small set of n-node graphs. Mutations can extend the graphs for high coverage. 
The framework applies these mutations by selecting different predefined mutation parameters, including 4 model-level mutations and 2 source code-level mutations.

%The framework applies mutations including 4 model-level mutations and 2 source-level mutations. The available mutations are the hyper parameters of mutation selector. With the hyper parameters set properly, models-level or source-level mutations can be produced. The mutation selector enumerates these mutations and the mutation selector can identity these mutations by their indices. These mutations are enumerated the mutation selector

Model-level mutations are applied to the initial digraph and blocks. Let ${E(g)}$ be the node count of graph ${g}$. Let ${r}$ ($0\leq{r}<1$) be the probability of model-level mutations.

\begin{itemize}
\item\textbf{Graph Edges Addition (GEA)}. Add $\lceil {E(g)}\cdot{r} \rceil$ edge to graph ${g}$.
\item\textbf{Graph Edges Removal (GER)}. Delete $\lfloor {E(g)}\cdot{r} \rfloor$ edge from graph ${g}$.
\item\textbf{Block Nodes Addition (BNA)}. Duplicate an operator to every subgraph of graph ${g}$ with probability ${r}$.
\item\textbf{Block Nodes Removal (BNR)}. Remove an operator and its edges from every subgraph of graph ${g}$ with probability ${r}$.
\end{itemize}

% \begin{itemize}
% \item\textbf{Edges of Graph Addition (EGD)}. Edges of the graph are duplicated with probability ${p}$ with the neighborhood count of each node not more than maximum neighborhoods ${k}$.
% \item\textbf{Edges of Graph Removal (EGR)}. Edges of the graph are deleted with probability ${p}$.
% \item\textbf{Edges of Block Re-connection (EBR)}. If a block is a subgraph, remove an edge and reconnect it to another node with each operator fulfill the input count limitation in corpus.
% \item\textbf{Nodes of Block Removal (NBR)}.If a block is a subgraph, remove an operator and its edges with  probability ${p}$. Reconnect the broken edges with each operator fulfill the input count limitation in block corpus.
% \end{itemize}

%\begin{table}[htbp]
%\centering
%\caption{Details of  Model-level Mutations Operators on DL inference engine}
%\label{table4}
%\begin{tabular}{c c c} %{|c|p{2.8cm}|c|p{2.6cm}|  }
%\hline
%Operator&Mutations Function&Description\\
%\hline
%EGD&${op_{egd} (edges,p)}$&Edges of Graph Duplication\\
%\hline
%EGR&${op_{egr} (edges,p)}$&Edges of Graph Removal\\
%\hline
%EBR&${op_{ebr} (edges,p)}$&Edges of Block Re-connection\\
%\hline
%NBR&${op_{nbr} (nodes,p)}$&Nodes of Block Removal\\
%\hline
%\end{tabular}
%\end{table}

Two source-level mutations mutate input shape of the network and operator parameters after blocks selected for nodes in digraph.

\begin{itemize}
\item\textbf{Tensor Shape Mutation (TSM)}. Mutate the shape of the input tensor. 
\item\textbf{Parameters Mutation (PM)}. Variation of input parameter. Selecting a random enumeration for a discrete type and a random value within range for continuous type.
\end{itemize}

\subsection{Input Mutator}

%\textbf{Input Mutator}. The input mutator mutates the input according to the selected blocks and mutations. Fig. \ref{p2} illustrates how the input mutation generator constructs mutated input test samples. 

To generate a DL model, the following steps are applied. First, network generator generates a digraph with a specific graph model and update connections of the graph by the model mutation methods. For each node in the graph, a block with the same input degree is selected from the block corpus.
Second, shapes\&parameters calculator calculates input shape and parameters for each input. Finally, generate models and test samples for running.

\textbf{Network Generator}. Two random graph models in graph theory are applied. One is Watts-Strogatz (WS) model proposed by Watts et.al.\cite{r49}, and the other is Residual Network (RN) model we proposed in this paper.
% and Barabasio-Albert\cite{r68}. And We found that using these two models will increase operator-level coverage faster than using one of the models.
% Residual Network model(RN) adds residual block to networks.
The RN model generates residual blocks in models. Let ${n}$ be the node count. Let ${k}$ (${k} \geq 2$) be the maximum neighbors. Initially, add the ${n}$ nodes ${i}=0,...,{N-1}$ sequentially in a line. Let ${k\_current_i}$ ($1 \leq {k\_current_i}\leq {k}$) be the neighbor count of node ${i}$. For every node whose current neighbor count is less than ${k}$, 
Add edges connecting a node ${i}$ to another node ${j}$ (${i} < {j}$ and ${k\_current_j} < {k}$) with probability ${p}$ ($0 < {p}\leq 1$), and repeat this step ${k}-{k\_current_i}$ times for node ${i}$. ${k}$, ${p}$ and ${n}$ are the parameters of the RN model, denoted as $RN({k}, {p},{n})$.

%\begin{table}[htbp]
%\centering
%\caption{Details of Source-level Mutations Operators on DL inference engine}
%\label{table5}
%\begin{tabular}{c c c } 
%\hline
%Operator&Mutations Function&Description\\
%\hline
%EGD&${op_{tsm}(network)}$&Tensor Shape Mutation\\
%\hline
%EGR&${{op}_{pm}(operator)}$&Parameters Mutation\\
%\hline
%\end{tabular}
%\end{table}

%\subsection{Shapes\&parameters Calculator}
\textbf{Shapes\&parameters Calculator}.
Shapes\&parameters calculator involves satisfying the demands of structuring DL neural network. Those shape-free parameters are randomly selected from their range. Two methods are used to focus on the effectiveness of DL models with various input shapes.
\begin{itemize}
\item\textbf{Aggregation},including Add, Concat, etc. 
We use operators such as Cast, Shape, Slice, and Pad to convert the mismatched input shape or data type into expected form before these aggregations.
Pad is merged into adjacent operators, such as Pooling, Conv2d, DepthwiseConv2d, Pad, etc.

\item\textbf{Operators with padding}, including Pooling, Conv2d, DepthwiseConv2d, etc. The padding of these operators are calculated to keep the shapes of input and output consistent. Taking Conv2d with 'SAME' padding mode for example, given input shape [iN,iH,iW,iC] and other parameters, the output height oH is computed as (\ref{eq_nhcw_1}), where ${pH}$  is padding height, ${fH}$ is filter height, ${sH}$ is stride height, and ${dH}$ is dilation height.

\begin{equation}
{oH=(iH+2\cdot pH-dH\cdot (fH-1))/sH}\label{eq_nhcw_1}
\end{equation}
Regarding the shapes of layer input and output are consistent, the following can be obtained: \begin{equation}
{oH = iH}\label{eq_nhcw_2}
\end{equation}
Then other parameters are associated, satisfying three conditions of Conv2d as below, where ${Max\_sH}$ is maximum of stride height, ${ Max\_dH}$ is maximum of dilation height, and ${fH}$ is maximum of ${pH}$. 
\begin{equation}
{0\leqslant pH \leqslant fH}\label{eq_nhcw_3}
\end{equation}
\begin{equation}
{1 \leqslant sH \leqslant Max\_sH}\label{eq_nhcw_4}
\end{equation}
\begin{equation}
{1 \leqslant dH \leqslant Max\_dH}\label{eq_nhcw_5}
\end{equation}
Similarly, parameters of filters can be computed in the same way. Thus with given input shape, parameters of input vector {[iN, iC, iH, iW, fN, fC, fH, fW, pad, stride, dilation]} that satisfy (\ref{eq_nhcw_1})-(\ref{eq_nhcw_5}), can be generated randomly with less Pads.

\end{itemize}

\section{Experiment Setup}
\subsection{Research questions}
In this paper, we will answer the following six research questions.

\textbf{RQ1}: How effective is the approach in detecting exceptions of DL inference engine? 

\textbf{RQ2}: How does the MCTS-based search algorithm perform comparing with random search for decision processes?

\textbf{RQ3}: How effective is the RN model in increasing operator-level coverage and detecting exceptions?  

\textbf{RQ4}: How effective is the mutation strategy in increasing operator-level coverage criterion and detecting exceptions?   

\textbf{RQ5}: How effective is the subgraph of the approach?  

\textbf{RQ6}: Are these exceptions found related to the operator-level coverage?

\subsection{Experiment setup}
We set up our experiments as follows.

\textbf{Block Corpus}. Block corpus in this experiment 
% are depicted as Table \ref{table_blockcorpus_setup} which
consists of 53 blocks, including 50 operators and 3 subgraphs. 

(1) 50 TensorFlow operators supported by MNN in reference guide\cite{r67}: Add, Argmax, Avgpooling, Batchtospacend, Biasadd, Cast, Ceil, Concat, Conv2d, Cropandresize, Deconv2d, Depthwise, Exp, Expanddims, Fill, Fusedbatchnorm, GatherV2, Greater, Lrn, Maximum, Maxpooling, Minimum, Mul, Realdiv, Reducemax, Reducemean, Reduceprod, Reducesum, Relu, Relu6, Reshape, Resizebilinear, Resizenearestneighbor, Rsqrt, Selu, Shape, Sigmoid, 
Resize, Slice, Softmax, Spacetobatchnd, Sqrt, Square, Squeeze, Stridedslice, Sub, Tanh, Tile, TopKV2, Transpose.

\begin{figure}[!ht]
\subfigure[Subgraph 1.]{
\includegraphics[width=0.19\textwidth]{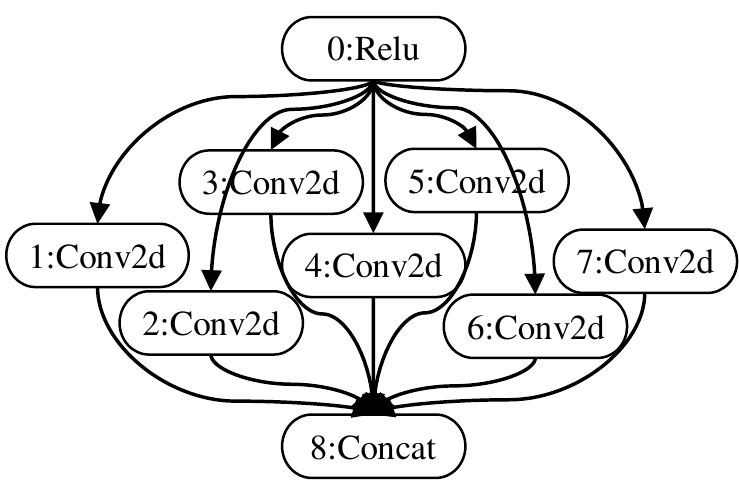}
\label{RQ4_subgraph_1}
}
\subfigure[Subgraph 2.]{
\includegraphics[width=0.13\textwidth]{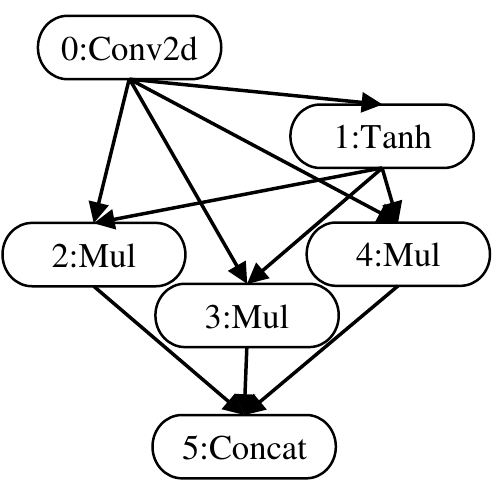}
\label{RQ4_subgraph_2}
}
\subfigure[Subgraph 3.]{
\includegraphics[width=0.10\textwidth]{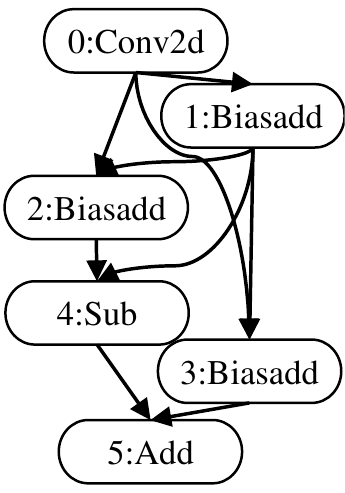}
\label{RQ4_subgraph_3}
}

\caption{Three subgraphs in block corpus
}
\label{RQ4_subgraph}
\end{figure}

(2) Subgraphs.  Subgraph 1 (Fig. \ref{RQ4_subgraph_1}) refers to the concatenation of multiple feature maps in SSD \cite{r12}.  Subgraph 2 (Fig. \ref{RQ4_subgraph_2}) is inspired by operator fusion and arithmetic optimizer in TensorFlow graph optimization. Subgraph 3 (Fig. \ref{RQ4_subgraph_3}) is inspired by operator fusion in MNN.
We set the range of output degree as \{0, 1, 2, 3, 4, 5\}. 

\textbf{Inference Runtime}. Our experiments are conducted with MNN (V1.0.2) on X86 CPU ( 2.10GHz $\times$ 4) and ARM CPU (Honor X10). Both inference runtimes are calculated using FP32. Test inputs and their inference results are generated by TensorFlow (V1.12, CPU mode). Inputs, filters and biases are generated from the uniform distribution [-1,1]. The final status of the test process includes model conversion failure (MCF), inference failure (IF), data comparison failure (DCF) and data comparison pass (DCP). 
The comparison threshold between MNN and TensorFlow is that the ratio of data with a relative error greater than 0.1\% to the total data is less than 0.1\%.

Formally, let ${RE}({mnn})$ be the ratio of the numbers with the relative error between MNN and TensorFlow over the total output data of an operator. 
% If ${RE}$ of a result is greater than 99.9\%, 
When ${RE}({mnn}) \geq 99.9\% $, the result is considered as a success.

Some failures are caused by the same defect. 
In order to eliminate duplication, model conversion failures with the same error type, error code and failure operator are regarded as the same error, and data comparison failures with the same structure and operator are regarded as duplicates.

%The probability of edges of Graph addition or removal is chosen from \{0, 0.05, 0.1, 0.3\}. The %probability of edges of blocks re-connection or removal is chosen from \{0, 0.05, 0.1, 0.3\}. 
\textbf{Configuration}. In order to generate diversified models, the parameters are set in a wide range. The probability of four models-level mutations is chosen from \{0, 0.1, 0.2\}. In graph algorithm, the number of neighbors $k$ is chosen from \{2, 4, 6\}. The probability of rewriting connections $p$  of WS is 0.5. The probability $p$  of RN is 0.9.
%The probability $p$  of RN is ultimately settled on the 0.9, which we found to work better.
For operator-level coverage, the ${n_{maxspc}}$ of Formula(\ref{eq_SPC}) is set to 200. The weights of Formula (\ref{eq_OLCOP}) and  (\ref{eq_OLC}) are set to 1.
% We make each strategy generate 400 inputs for RQ2, RQ3 and RQ4, and each strategy runs 5 times to get 5 sample coverage and inference results. 
% The block number of a model is set as 5, 10 and 15.
In MCTS-based block chooser, ${tc1}$ is set to 10, ${tc2}$ is set to 1 and $e$ is set to $1/\sqrt{2}$. 
Each terminal node can expand up to 3 child nodes.

For RQ2, RQ3 and RQ4, 400 test inputs are generated for each strategy, and they are inferred on x86 CPU only, because MNN cannot infer models with multiple inputs or multiple outputs on ARM CPU.
The block number of models for each strategy is set to 5, 10 and 15 respectively.
For RQ1 and RQ5, the generated models are inferred on X86 CPU,and the single-input and single-output models of these models are also inferred on the ARM CPU.

\section{Evaluation}
\subsection{RQ1: How effective is the approach in detecting exceptions of DL inference engine?  }

In order to answer RQ1, we generated approximately 1000 models and found more than 40 different exceptions. The number of blocks for models is chosen uniformly from 1 to 30. Some typical exceptions are as below. 

\textbf{Models Conversion Failure (MCF)}.

\textbf{MCF-1 Segmentation fault in TopKV2 conversion.} MNN model cannot be generated. MNN Log: Error for 18. Segmentation fault (core dumped). Almost all models that include TopKV2 operators cannot be converted and inferred successfully.

\begin{figure}[ht]
 \centering
 \includegraphics[width=0.38\textwidth]{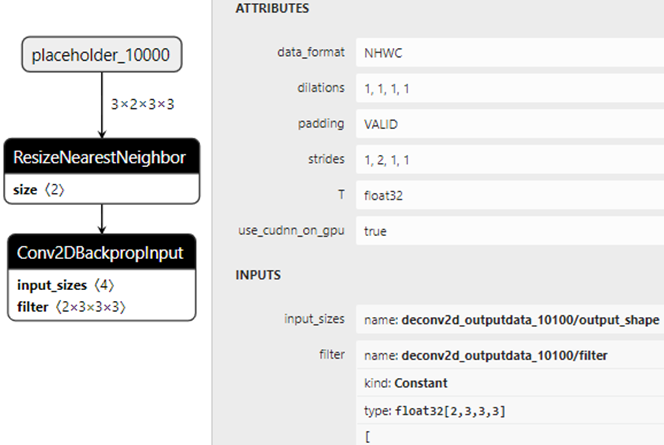}
 \caption{MCF-2 Conversion aborted: pb model structure and deconv‘s parameters.}
 \label{RQ1_MCF_2}
\end{figure}

\textbf{MCF-2 Conversion aborted.} The deconv API of the pb model(shown in Fig. \ref{RQ1_MCF_2}) is tf.nn.conv2d\_transpose. MNN log: /converter/source/common/writeFb.cpp:108: Check failed: (notSupportOps.size()) == (0). Not Support: Tensorflow::Conv2DBackpropInput, output\_shape is not consistent with inferred output shape in MNN. (height, width): (38,102) vs (4,21). Convert Tensorflow's Op deconv2d\_outputdata\_10100, type = Conv2DBackpropInput, failed. Aborted (core dumped). It reveals that the shape of the output tensor calculated by the operator Deconv in MNN is wrong.

\begin{figure}[ht]
 \centering
 \includegraphics[width=0.45\textwidth]{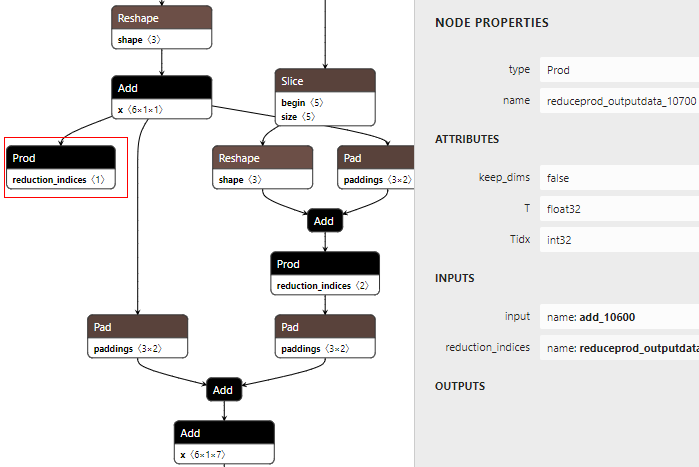}
 \caption{IF-1 Inference aborted: pb model structure and Reduceprod‘s parameters. Inference aborted: the reduceprod cannot output result.}
 \label{RQ1_IF_1}
\end{figure}

\textbf{Inference Failure(IF)}.
\textbf{IF-1 Inference aborted}. The Reduceprod cannot output results (shown in Fig. \ref{RQ1_IF_1}). MNN Log: Error in `python': free(): invalid next size (fast): 0x000000000  1e2cb90. Backtrace: .../dist-packages/\_mnncengine.so (\_ZN3MNN6TensorD1Ev+0x74) [0x7f7672028654]. Aborted (core dumped).

\begin{figure}[ht]
  \centering
  \includegraphics[width=0.4\textwidth]{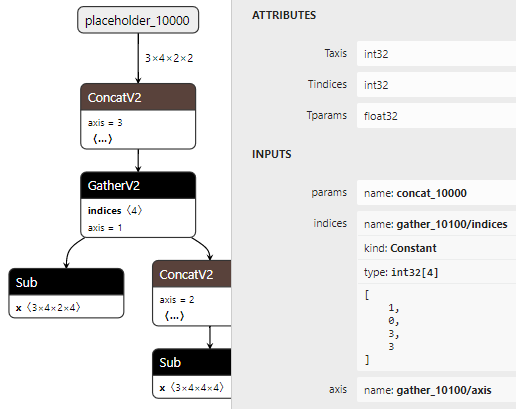}
  \caption{DCF-1: pb model structure and Gather‘s parameters. Data Comparison Failure of Sub.}
  \label{RQ1_DCF_1}
\end{figure}

\textbf{Data Comparison Failure(DCF)}
\textbf{DCF-1 core dumped and data comparison failure in Sub}. The model structure is shown in Fig. \ref{RQ1_DCF_1}. ${RE}({mnn_{X86CPU}})$ of Sub1 is 52.08\%, ${RE}({mnn_{X86CPU}})$ of Sub2 is 76.04\%.
 MNN log: Error in `python': double free or corruption (!prev): 0x0000000002 4ae8b0. Backtrace: /dist-packages/\_mnncengine.so (\_ZN3MNN15BufferAllocator4NodeD1Ev+0x88) [0x7ff0e81 7a098]. Aborted (core dumped).    

\begin{figure}[ht]
\subfigure[${RE}({mnn_{ARMCPU}})$ of Avgpooling is 33.33\%.
% The relative errors between avgpooling's results of MNN(Android) and TensorFlow is 33.33\%.
]{
\includegraphics[width=0.21\textwidth]{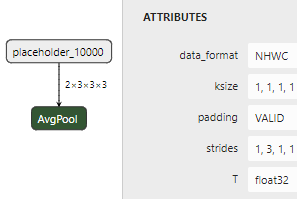}
\label{RQ1_DCF_2_a}
}
\subfigure[${RE}({mnn_{ARMCPU}})$ of Maxpooling is 50.00\%.]{
\includegraphics[width=0.21\textwidth]{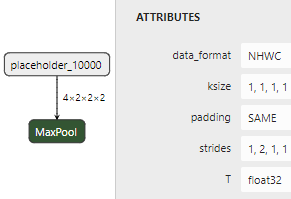}
\label{RQ1_DCF_2_b}
}
\subfigure[${RE}({mnn_{ARMCPU}})$ of Relu is 45.19\%.]{
\includegraphics[width=0.24\textwidth]{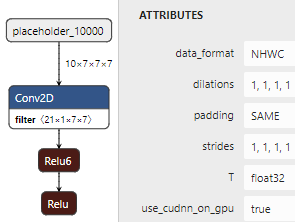}
\label{RQ1_DCF_3_c}
}
\subfigure[${RE}({mnn_{ARMCPU}})$ of Realdiv is 91.67\%.]{
\includegraphics[width=0.20\textwidth]{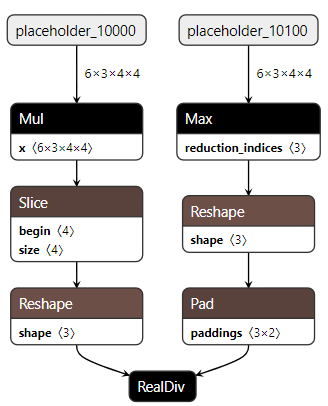}
\label{RQ1_DCF_3_d}
}

\caption{DCF-2 Data Comparison Failure of single operators 
}
\label{RQ2}
\end{figure}

\textbf{DCF-2 Data Comparison Failure of operators}. (1) When converting some FP32 numbers(from 0 to 1) into INT8 numbers, the cast operator will yield a calculation error. (2) When the input of a sigmoid operator is nan (coming from Rsqrt), the results are nan in TensorFlow and 1 in MNN respectively.
(3) In Fig. \ref{RQ1_DCF_2_a}, ${RE}({mnn_{ARMCPU}})$ of Avgpooling is 33.33\%.
% The relative errors between avgpooling's results of MNN (ARM CPU) and TensorFlow is 33.33\% (Figure \ref{RQ1_DCF_2_a}).
(4) In Fig. \ref{RQ1_DCF_2_b}, ${RE}({mnn_{ARMCPU}})$ of Maxpooling is 50.00\%.
% The relative errors between maxpooling's results of MNN (ARM CPU) and TensorFlow is 50.00\% (Figure \ref{RQ1_DCF_2_b}).
% When the n of filter is not 1, deconv operator and depthwise operator will make a calculation error. 
(5) In Fig. \ref{RQ1_DCF_3_c}, ${RE}({mnn_{ARMCPU}})$ of Relu is 45.19\%.
% ${RE}$ between the results  of MNN (ARM CPU) and TensorFlow is 45.19\%.
The Conv2d of the model (Fig. \ref{RQ1_DCF_3_c}) caused an incorrect calculation result.
(6) The result of Realdiv (Fig. \ref{RQ1_DCF_3_d}) divided by zero is inf in TensorFlow and nan in MNN X86 CPU.

\textbf{Model Generation Failure}.
It is worthy of mentioning that some exceptions of TensorFlow are found in model generation. When generating a model containing two Concats whose two inputs come from the same two constants, TensorFlow will get stuck.

\textbf{Answer to RQ1}: %Guided by operator-level coverage, 
Our approach is effective to detect various exceptions in model conversion and inference process for the DL inference engine, such as crashes, inconsistencies, nan and inf bugs.

\subsection{RQ2: How does the MCTS-based search algorithm perform comparing with random search for decision processes?}
In order to answer RQ2, we evaluate operator-level coverage and inference results using MCTS-based search and random search for block chooser. %We make each strategy generate 400 input samples with blocks that supported by SNPE in block corpus.
% The block number of each neural network is generated from the uniform distribution [5, 20]. In MCTS-based block chooser, ${tc1}$ is set to 10, ${tc2}$ is set to 1 and e is set to $1/\sqrt{2}$.

%For each block number, we run 10 times to get 10 sample coverage and inference results.

%Different from existing fuzz testing, our method generates models as the input of the fuzz testing for DL inference engine. 

\textbf{Operator-level coverage}. 
% The operator-level coverage of two search algorithms are 90.6\% and 87.1\% respectively, and MCTS-based search, on average, covers 3.5\% more operator-level coverage than random search where number of test sample is 400 as demonstrated in Table \ref{table_RQ234}.
% Figure .
As shown in TABLE \ref{table_RQ234} (1)(5), the operator-level coverage of random search and MCTS-based search are 60.2\% and 61.5\% for 5 blocks, 69.9\% and 76.4\% for 10 blocks, 71.8\% and 81.7\% for 15 blocks respectively. MCTS-based search, on average, covers 6.7\% more operator-level coverage than random search.
% where number of test sample is 400 as demonstrated in Table \ref{table_RQ234}.

\textbf{Inference results}. We measure the exceptions of inference results for each search algorithm. 
TABLE \ref{table_RQ234}(1)(5) shows the exceptions found. MCTS-based search, on average, finds 9.7 more exceptions than random search.
It can be observed that the average unduplicated rate of random search is 11.6\%, which is lower than that of MCTS search (28.3\%). %As shown in Table III, our method(Line 2, 3, 4) finds more valid exceptions(Column Exceptions found).
%random search, on average, finds 53.3 more repeated exceptions than MCTS-based search. 
Using the MCTS-based search takes an average of 1.1 hours longer than using the random search. Because the efficiency of  MCTS-based search that find exceptions is significant, a small increase in execution time is acceptable for industrial testing.

\textbf{Answer to RQ2}: The MCTS-based block chooser outperforms the random-based block chooser in boosting operator-level coverage (6.7\% more) and detecting exceptions (9.7 more).
%Through applying the MCTS-based search to block chooser, MCTS-based search can detect more exceptions earlier than random search as well as operator-level coverage increasing. 

\begin{table*}[h!]
\caption{Operator-level coverage, exceptions found (after deduplication/total) and duration under different block number (n = 5, 10, 15) of models for RQ2, RQ3 and RQ4.}
\scriptsize
\label{table_RQ234}  
%\begin{tabular}{|p{0.58cm}<{\centering}|p{3.2cm}<{\centering}|p{1cm}<{\centering}|m{2.2cm}<{\centering}|}
\begin{tabular}{m{0.3cm}<{\centering}m{1.5cm}<{\centering}m{0.2cm}<{\centering}m{1.5cm}<{\centering}m{0.8cm}<{\centering}m{1.3cm}<{\centering}m{0.9cm}<{\centering}m{0.8cm}<{\centering}m{1.3cm}<{\centering}m{0.9cm}<{\centering}m{0.8cm}<{\centering}m{1.3cm}<{\centering}m{1.2cm}<{\centering}}
 \hline

 &  &  &  &N = 5& & &N = 10& & &N = 15 & &   \\
 \hline
  & Graph model & Mutations & Search & OLC&Exceptions found (d/t) &Duration (h) & OLC & Exceptions found (d/t) & Duration (h) & OLC & Exceptions found (d/t) & Duration(h)\\
 \hline
1 & WS and RN & Yes & Random & 60.2\% &12/103 &1.6 &  69.9\% & 20.1/161 &2.7 &  71.8 \% & 21.7/201& 3.4\\
\hline
2 & WS  & NO & MCTS-based&   57.2\% &10.9/75.3  &2.7 &  62.9\% & 13.7/95.4 &4.1 &  64.8\% & 14.2/91.8& 4.4\\
\hline

3 & RN  & NO & MCTS-based&        57.7\% &14.4/77.2  &2.6 &  65.3\% & 19.5/92.8 &4.1 &  67.3\% & 17.9/113.4& 4.5\\
\hline

4 & WS and RN & NO & MCTS-based&  59.2\% & 16.7/81.2 & 2.8&  65.8\% & 20.2/93.1 &4.1 &  69.7\% & 19.5/118.9& 4.5\\
\hline
5 & WS and RN & Yes & MCTS-based&  61.5\% & 20.1/84.5 & 2.7&  76.4\%  &  31.9/89.6& 4.1&  81.7\% & 31.2/120.3& 4.5\\
\hline

\end{tabular}
\end{table*}

\subsection{RQ3: How effective is the RN model in increasing operator-level coverage and detecting exceptions?}

In order to answer RQ3, operator-level coverage and inference results are evaluated using two stochastic network generation strategies: (1) WS and RN model together, each model generating half of the test samples. We do not aim to verify that whether a graph model is superior to other models in certain configurations.
The RN model is only used to generate more diversified models. (2) WS model only. (3) RN model only. 
Five input shapes can be chosen uniformly for each model. To avoid disturbing the coverage and inference result of random graph models, mutations are cancelled. 
% And the block number of each neural network is set to 5, 10 and 15 respectively.
%We make each strategy generate 400 input samples with blocks that supported by SNPE. 
%and the block number of each neural network is set to 5, 10 and 15 respectively. For each block number, we run 10 times to get 10 sample coverage and inference results.

\textbf{Operator-level coverage}. As shown in TABLE \ref{table_RQ234}(2)(3)(4), 
%the operator-level coverage of two graph model are 57.2\% and 59.2\% for 5 blocks, 62.9\% and 65.8\% for 10 blocks, 64.8\% and 69.7\% for 15 blocks respectively. 
there are two key observations from the results. First, WS and RN model together, on average, covers 3.6\% and 1.47\% more operator-level coverage than WS model only and RN model only respectively as demonstrated. Second, 
the coverage rate when n is 15 is slightly higher than when n is 10.
% as the block number increases, the two strategies cover more.
This is intuitive as a higher value (N = 15) of blocks making it increasingly harder to cover more topologies and types of shapes\&parameters without mutations.

\textbf{Inference results}. The exceptions of inference results are measured for each strategy. TABLE \ref{table_RQ234}(2)(3)(4) shows the detailed results. 
%The mean number of exceptions are 10.9 and 16.7 for 5 blocks, 13.7 and 20.2 for 10 blocks, 14.2 and 19.5 for 15 blocks respectively. 
 WS and RN model together, on average, finds 5.9 and 1.5 more exceptions than WS model only and RN model only respectively as demonstrated.

Therefore WS and RN model together is more efficient in finding exceptions as well as increasing blocks of DL models. The experiment also shown that the performance of RN is slightly better than that of WS.

\textbf{Answer to RQ3}: 
Different graph models can be used to generate more diverse models. Through applying the RN model to stochastic network generation strategy, more exceptions for each strategy can be detected as well as operator-level coverage increased.

\subsection{RQ4: How effective is the mutation strategy in increasing operator-level coverage criterion and detecting exceptions? }

In order to answer RQ4, we evaluate operator-level coverage and inference results of test generation with mutations and test generation without mutations. For test generation without mutations, 5 input shapes can be chosen uniformly for each model without any mutations. 
% The block number of neural network is set as 5, 10 and 15.

\textbf{Operator-level coverage}. The operator-level coverage of two strategies are 59.2\% and 61.5\% for 5 blocks, 65.8\% and 76.4\% for 10 blocks, 69.8\% and 81.7\% for 15 blocks respectively. For test generation with mutations, on average, covers 8.2\% more operator-level coverage than test generation without mutations as demonstrated in TABLE \ref{table_RQ234}(4)(5).
% Figure \ref{RQ2_olc}.

\textbf{Inference results}. The exceptions of inference results are measured for each strategy. 
TABLE \ref{table_RQ234}(4)(5)
shows the detailed results. The mean number of exceptions are 16.7 and 20.1 for 5 blocks, 20.2 and 31.9 for 10 blocks, 19.5 and 31.2 for 15 blocks respectively. It can be observed that test generation with mutations is more efficient in finding exceptions(on average, 8.6 more exceptions) as well as increasing operator-level coverage of inputs.

\textbf{Answer to RQ4}: Our mutation strategy is useful to generate new valid test inputs, by up to 8.2\% more operator-level coverage and 8.6 more exceptions detected on average.
It can be observed that our mutation strategy is effective for generating diversified models.

\subsection{RQ5: How effective is the subgraph of the approach?
}

\begin{figure}[!ht]

\subfigure[Delete a Mul from subgraph 2.   
% DCF and RE = 37.96\%
]{
\includegraphics[width=0.11\textwidth]{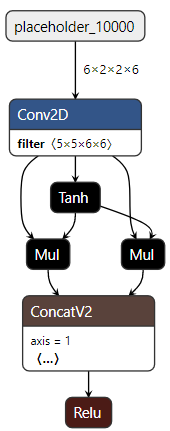}
\label{MS_1_1}
}
\subfigure[Delete a Biasadd from subgraph 3, and delete a Conv2d from subgraph 1.
% DCF and RE = 8.93\% 
]{
\includegraphics[width=0.34\textwidth]{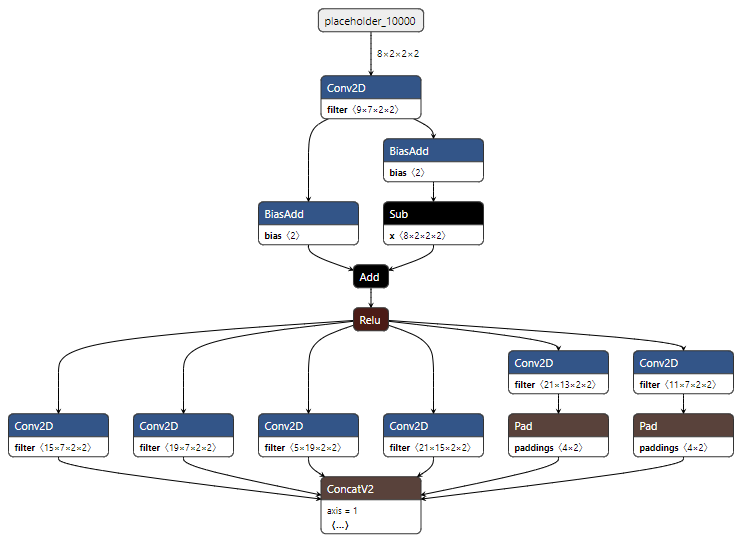}
\label{MS_1_2}
}
\caption{Inference results of 3 mutated subgraphs in block corpus.}
\label{RQ5_MS_1}
\end{figure}

In order to answer RQ5, 100 mutated subgraphs (MS) are generated to evaluate the effectiveness of three subgraphs in block corpus (shown in Fig. \ref{RQ4_subgraph}). The number of blocks in a model is chosen from \{1, 3\}. The success ratios are 45.2\% on ARM CPU and 100\% on X86 CPU. The main type of error is data comparison failure on ARM CPU. Inference exceptions are analyzed as below.
Fig. \ref{RQ5_MS_1} shows two typical exceptions of mutated subgraphs. 
% The success ratio is 45.2\% on ARM. 
%  And these models run on X86 are all passed. 

\textbf{MS-1}. 
% The success rate of block inference result is 20\%. The major error types are Inference Failure (error\_code=910 only) and Data Comparison Failure. Figure \ref{RQ5_MS_1} shows two typical mutated subgraph.  
Data Comparison Failure. This mutant deletes a Mul from subgraph 2  (Fig. \ref{MS_1_1}) and ${RE}({mnn_{ARMCPU}})$ of Relu is 33.33\%.

\textbf{MS-2}. 
Data Comparison Failure. This mutant a Biasadd from subgraph 3, and delete a Conv2d from subgraph 1 (Fig. \ref{MS_1_2}) and ${RE}({mnn_{ARMCPU}})$ of Concat is 8.93\%. when given more than 2 inputs, it reveals that the operator Concat will lose some of the input data and outputs wrong results.

% The success rate of block inference result is 52\%. The major error types are also Inference Failure (DSP only, error\_code=910) and Data Comparison Failure (both DSP and CPU runtime). Figure \ref{RQ5_MS_2}  shows two typical mutated subgraph and their inference result that run failed in CPU runtime or DSP runtime.  

% \begin{figure}[!ht]
% \subfigure[Delete Relu, Data Comparison Failure of Concat on DSP (35.08\%)]{
% \includegraphics[width=0.17\textwidth]{pictures/RQ5/MS_2_1.png}
% %\includegraphics[width=0.7\linewidth]{pictures/RQ5/MS_2_1.png}
% \label{RQ5_MS_2_1}
% }
% \subfigure[MS-2 Delete a Conv2d and reconnect an edge between Relu and Conv2d. Data comparison failure of Concat on DSP (58.80\%).]{
% \includegraphics[width=0.22\textwidth]{pictures/RQ5/MS_2_3.png}
% %\includegraphics[width=0.7\linewidth]{pictures/RQ5/MS_2_3.png}
% \label{RQ5_MS_2_3}
% }
% %\subfigure[Delete a Concat, Data comparison failure on CPU (0.69\%, 0.69\%, 0.00\%, %0.35\%,0.69\%,0.87\%).]{
% %\includegraphics[width=0.18\textwidth]{pictures/RQ5/MS_2_2.png}
% %\includegraphics[width=0.7\linewidth]{pictures/RQ5/MS_2_2.png}
% %\label{RQ5_MS_2_2}
% %}

% \caption{Inference result of mutated subgraph of block 22 in block corpus}
% \label{RQ5_MS_2}
% \end{figure}
\textbf{Answer to RQ5}: It is difficult to generate a model of a specific structure only by matching operators with nodes in a random graph.
% Subgraphs defined in block corpus could increase the probability of occurrence of specific structures.
Subgraphs and their mutants can be helpful to construct these specific structures with more exceptions detected. 
% Mutants of these subgraphs  needed to enhance diversity of models.

\subsection{RQ6:  Are these exceptions found related to the operator-level coverage?
}
In order to answer RQ6, relations between the typical exceptions and operator-level coverage are analyzed.

\textbf{Operator Type Coverage}. We note that some operators are supported in MNN reference guide,
but the error reported is that this operator is not supported in model conversation, such as Deconv in MCF-2. Almost all models that include TopKV2 operators cannot be converted and inferred successfully. Our result(MCF-1) confirm that the TopKV2 operator is almost unsupported in current MNN version.
We also tried to infer some models that contain operators unsupported in MNN reference guide, such as Addn, Clip and Asin. These models are successfully inferred by MNN on X86 CPU and ARM CPU.

% Taking MCF-2 for example, Deconv is supported in MNN reference guide, but the error reported in log is that this operator is not supported. 

\textbf{Single Edge Coverage}. Note that
% Take MCF-2 for example, Deconv can be converted correctly in a single operator model, but the error reported in log is that this operator is not supported when It is connected after ResizenearestNeighbor. 
SEC is usually associated with \textbf{Input Degree Coverage} or \textbf{Output Degree Coverage}, such as multi-output (e.g., add of IF-1) and multi-input operators (e.g., (6) Realdiv of DCF-2 and Concat of MS-2). Some structures also cause special inputs for operators, such as nan of (2) Rsqrt and Sigmoid in DCF-2.

% \textbf{Multiple Edges Coverage}  is usually associated with Input degree of Operator Coverage, such as MCF-1. In-degree of multi-input operators (i.e. Add, Addn, Concat, etc) in SNPE are different with TensorFlow.

\textbf{Shapes\&Parameters Coverage}. Some results of operators with specific value of parameters or tensor shapes are unexpected in comparison with TensorFlow, such as Avgpooling and Maxpooling in DCF-2.

\textbf{Answer to RQ6}: In summary, the exceptions detected are all within the scope of operator-level coverage. 
In addition, there is no obvious correlation between each metrics of operator-level coverage and a certain error types. That is, guided by these metrics, the inputs can trigger various types of exceptions.

\section{Threats To Validity}
The internal threat to validity mainly lies in the implementations of our method and scripts in the experiment. All the artifacts are maintained by experienced software engineers in Huawei. Besides, the method has been used in Huawei company, which reduces the internal threats. 

The external threats to validity mainly lie in the  library and DL inference engine used in the experiment. To reduce this threat, we adopted one of the most widely-used library TensorFlow to generate input DL models. Then, we chose the Alibaba MNN as the target DL inference engine under test (X86 CPU and ARM CPU). Furthermore, the proposed method tests a DL inference engine of Huawei for several months with many valid exceptions detected. 

The construct threats to validity mainly lie in  settings, randomness and measurements in our experiment. (1) To reduce the impact of settings, we constructed a block corpus (50 operators and 3 subgraphs ) and five experiments. (2) To reduce the impact of randomness, a large number of models were generated in every experiment respectively (i.e., 1000 for RQ1, 400 for RQ2-RQ4). We repeated each method for 10 times and calculated the average results in RQ2-RQ4. (3) To reduce the impact of measurements, we carefully set a threshold ${RE}({mnn})$ to check whether the inference results are correct. When counting the number of exceptions, we identified the duplicated ones. Further, we manually generated DL models and triggered all the  detected exceptions successfully.

\section{Related Work}
\textbf{Fuzz Testing and Mutation Testing}. Fuzz testing is a widely used technique for exposing defects in DL system. Guo et al. \cite{r43} proposed the first differential fuzz testing framework for DL systems. TensorFuzz proposed by Odena et al. \cite{r1}, used a nearest neighbour hill climbing approach to explore achievable coverage over valid input space for TensorFlow graphs, and to discover numerical errors, disagreements between DL models and their quantized versions. Pei et al. presented DeepXplore \cite{r2} which proposed a white-box differential testing technique to generate test inputs for DL system. Wicker et al. \cite{r41} proposed feature-guided test generation. They transformed the problem of finding adversarial examples into a two-player turn-based stochastic game. Ma et al. \cite{r9} proposed Deepmutation which mutates DNNs at the source level or model level to make minor perturbation on the decision boundary of a DNN. Shen et al. \cite{r36} proposed five mutation operators for DNNs and evaluated properties of mutation . Xie et al. \cite{r45} presented a metamorphic transformation based coverage guided fuzzing technique, DeepHunter, which leverages both neuron coverage and coverage criteria presented by DeepGauge \cite{r46}.
Existing testing techniques focus on the quality of
DL models but lacks attention to the core underlying inference
engines (i.e., frameworks and libraries). Our method generates models as the input of the fuzz testing for DL inference engine. Together with the combinations of operators, we design new mutation rules to generate diversified DL models to trigger different structured parts of a given DL inference engine.

\textbf{Test Coverage Criteria}. Coverage criteria is an important part of testing methods. Code coverage is most popular for conventional software testing, but it does not make sense for DL testing, since the decision logic of a DL model is not written manually but rather it is learned from training data \cite{r42}. In the study \cite{r2}, 100 \% traditional code coverage is easily achieved by a single randomly chosen input. Pei et al. \cite{r2} proposed the first Coverage criterion: neuron Coverage. Neuron coverage is calculated as the ratio of the number of unique neurons activated by all test inputs and the total number of neurons. Ma et al. \cite{r35} proposed layer-level Coverage, which considers the top hyperactive neurons and their combinations to characterise the behaviours of a DNN. Du et al. \cite{r36} first proposed State-level Coverage to capture the dynamic state transition behaviours of deep neural network. Li et al. \cite{r38} pointed out the limitations of structural coverage criteria for deep networks caused by the fundamental differences between neural networks and human-written programs. DeepCover \cite{r53} proposes the test criterion \cite{r54} for DNNs, adapted from the MC/DC test criterion of traditional software.
Existing neural coverage criteria of DL models cannot work in DL inference engine testing scenario, because the inputs for testing DL inference engines are DL models. 
Thus, a novel criterion is required to capture behaviors of DL inference engines rather than those of DL models. 
Our proposed operator-level coverage is naturally followed by the graph structure of DL models. The results show that the operator-level coverage guided testing framework improves the effectiveness in detecting exceptions.

%Their initial experiments with natural inputs found no strong correlation between the number of %misclassified inputs in a test set and its structural coverage.

% \textbf{Neural Architecture Search (NAS)}. Zoph et al. \cite{r14} define a NAS search space and investigate reinforcement learning (RL) as an optimization algorithm. Li et al. \cite{r39} demonstrate that random search with weight-sharing can outperform a series of powerful methods. Xie et al. \cite{r4} further  use three classical random graph models to generate randomly wired graphs for networks. Several variants of these random generators yield network instances that have competitive accuracy on the ImageNet benchmark.

\section{Conclusion}

The issues triggered by a single specific model are limited and diversified combinations of operators (models) are more capable of triggering DL inference engine issues. %We need to generate a large number of models by combining operators. 
To overcome the challenge of generating such models with diversified combinations of operators, this paper employs graph-based fuzz testing incorporating six different mutations, MCTS and a novel operator-level coverage criterion proposed based on graph theory. 
One possible application scenario is implemented on MNN, and the results demonstrate the effectiveness of our proposed method.
% More than 40 different exceptions have been discovered and categorized in three types of undesired behaviors: model conversion failure, inference failure, output comparison failure.
In fact, the proposed method has been used continuously in a DL inference engine of Huawei for several months, which finds many valid exceptions during the internal build and is efficient for industry.

\section*{Acknowledgement}
We would like to thank anonymous reviewers for insightful comments; we also thank Jiawei Liu for discussions on the manuscript. This work is supported partially by National Natural Science Foundation of China (61932012, 61802171), and Fundamental Research Funds for the Central Universities (14380021).

%This method is designed following by the graph structure of DL systems with some elements of neural network properties. An operator-level coverage based on graph theory is introduced and six different mutations are implemented to generate diversified DL models by exploring combinations of model structures, parameters and data.  MCTS is used to drive DL model generation without a training process.  The experimental results show that the MCTS outperforms random method in boosting operator-level coverage and  detecting exceptions.  Our method has discovered more than 40 different exceptions in three types of undesired behaviors: model conversion failure, inference failure, output comparison failure.  The mutation strategies are useful to generate new valid test inputs, by up to a 8.2\% more operator-level coverage on average and 8.6 more exceptions captured. 

% More details of graph-based fuzz testing and the experiments can be found at https://github.com/gbftdlie/Graph-based-fuzz-testing.
%The fuzzer can be used to measure test adequacy for SoC manufacturers, and also be used to select %DL inference and NNs design constraints for developers.

%This also indicates that new efforts focusing on designing NN generators with may lead to new %breakthroughs by exploring less constrained topologies or parameters of operators in input spaces.

%%
%% The next two lines define the bibliography style to be used, and
%% the bibliography file.
% \bibliographystyle{ACM-Reference-Format}
\bibliographystyle{IEEEtran}
\bibliography{reference}

\end{document}